\def\det{{\rm det}}

\documentclass[%
 reprint,
nofootinbib,
 amsmath,amssymb,
 aps,
prstab,
]{revtex4-2}

\usepackage{xcolor}

\usepackage{placeins}
\usepackage{graphicx}
\usepackage{dcolumn}
\usepackage{bm}
\usepackage{hyperref}
\usepackage{natbib}

\usepackage{subcaption}
\usepackage{caption}
\DeclareMathOperator\erf{erf}

\begin{document}

\title{Propagation of a Gaussian Wigner Function Through
         a Matrix-Aperture Beamline}

\author{Ilya V. Pogorelov}%
\email{ilya@radiasoft.net} 
\email{ilya.v.3.14159@gmail.com}
\author{Boaz Nash}%
 \email{bnash@radiasoft.net}
 \homepage{}
\author{Dan T. Abell}
\author{Paul Moeller}
\author{Nicholas Goldring}
\affiliation{RadiaSoft LLC, 6525 Gunpark Dr, Boulder CO 80301}
 \altaffiliation[]{}

\date{\today}

\begin{abstract}
In the framework of statistical optics, a Wigner function represents partially coherent radiation. 
A Gaussian Wigner function, which is an equivalent representation of the more commonly used Gaussian Schell-model cross-spectral density, may be defined in terms of its covariance matrix and centroid.  
Starting from the relationship between Gaussian Wigner functions and the Gaussian Schell model, we derive coherence properties of the Gaussian Wigner function, including coherence length and degree of coherence. 
We define a simplified beamline called a matrix-aperture beamline composed of linear transport sections separated by physical apertures. 
This is an idealized form for a transport beamline in a synchrotron light source or X-ray free electron laser. 
An envelope model provides a basic foundation for understanding the optics of a given beamline, in a manner analogous to how linear optics are treated in particle beam dynamics, with corresponding definitions of emittance and Twiss parameters. 
One major challenge to such an envelope model lies in the hard-edge apertures which break the Gaussian condition, raising the question as to the adequacy of a Gaussian model. 
We present a consistent way to construct a Gaussian approximation of the far-field Wigner function following the hard edge aperture. 
To this end, we introduce the concept of a Gaussian aperture and analyze its effects on the radiation Wigner function. 
A software implementation of this model is described as well.
\end{abstract}

\maketitle


\section{Introduction}
\label{sec:01}
X-ray beamlines are crucial components of synchrotron light sources, transporting the X-rays from the source to the sample. Whereas modeling of the X-ray transport is typically done during the design phase of a new beamline, it is less often carried out on an operating beamline. We suppose that part of the reason for this, is a gap in the basic linear optics formalism that would allow for rapid, yet accurate calculation of photon beam properties along the beamline.  \footnote{For one attempt in this direction, see \cite{Ferrero2008Aug} by C. Ferrerro et. al. Note however, that there is a problem with the authors' treatment of the impact of apertures on the Wigner function, which does not take into account the angular convolution. }

In this work, we will fill this gap by developing a Gaussian model for radiation transport that includes the effects of both linear transport and physical apertures, which are needed on most all beamlines to reduce the beam size, increase the stability and increase the coherence of the transported radiation.

\begin{figure*}[!tbh]
\begin{center}
\includegraphics[width=\textwidth,height=3.2in,keepaspectratio]{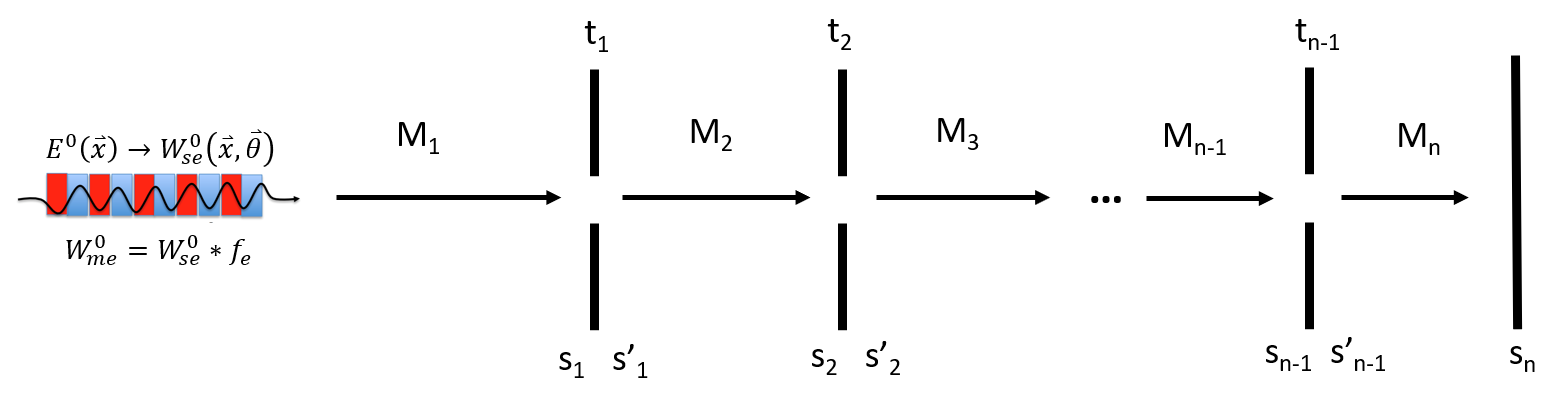}
\end{center}
\caption{Matrix aperture beamline schematic. We assume a source coming from an electron beam
passing through an undulator. The multi-electron (\textit{me}) Wigner function is then given as a 
convolution of the single electron (\textit{se}) Wigner function with the electron beam distribution $f_e$.}
\label{beamlineSchematic01}
\end{figure*}

As a general framework for the reduced model developed in this paper, we begin with what we call a {\it matrix-aperture beamline} illustrated in Figure \ref{beamlineSchematic01}. The radiation source is created by an electron beam passing through a magnetic field such as a bending magnet or undulator. The simplified beamline is then represented by a series of linear transport sections with ABCD matrices $M_1$ through $M_n$. A series of physical apertures are also included with transfer functions $t_1$ through $t_{n-1}$.

The structure of the paper is as follows. In section II, we describe the meaning of the Gaussian Wigner function and what can be computed from it, partly by means of the mapping to  the Gaussian Schell model (GSM). We parameterize the Gaussian beam using notation drawn from the particle beam literature, so that the theory can be grasped as easily as possible for those extending treatment of particle beam dynamics into the synchrotron radiation domain. 

In Section III, we consider the hard-edge aperture case. We derive the flux reduction formula for a Gaussian Wigner function. We introduce the concept of an aperture ``Wigner function" and derive a general analytical result for the Wigner function following passage through the hard-edge aperture.

Next, in Section IV, we consider the concept of a Gaussian aperture transfer function, for which the Wigner function remains Gaussian throughout.   We use the Gaussian aperture model to derive a number of analytical relations for the Wigner function propagation through the aperture. One notable result is that the transverse coherence length does not change when passing through a Gaussian aperture, although other coherence properties are affected and the coherence length will be affected further downstream. 

In Section V, we combine the results of the previous two sections to show how one can find a Gaussian approximation for the Wigner function following the aperture in the far-field regime.
Further, we find parameters for a Gaussian aperture approximation to the hard-edge aperture. These parameters are to be chosen such that the covariance matrix of the resulting Gaussian will match the hard-edge result in terms of full width at half maximum (FWHM) in the far-field. We provide numerical examples indicating that in practice, a Gaussian approximation is actually quite adequate at a modest distance from the aperture.

Finally, in Section VI, we describe the implementation of the Gaussian Wigner function method within the context of a realistic X-ray optics modeling code, SHADOW\cite{SanchezdelRio_2011}. The ray tracing software is used to compute transfer matrices propagating the covariance matrix. This has been implemented in the Sirepo interface to the SHADOW code\cite{SirepoShadow}, and some results for a simple beamline with a Kirkpatrick-Baez (KB) mirror focusing element and a physical aperture are presented here. Section VII contains our summary and conclusion, followed by several appendices where additional details of the calculations are presented.

\section{Gaussian Wigner Function Model for Partially Coherent X-Ray Beams}
\label{sec:02}

\subsection{Correlation Functions of the Radiation Field}


Within the statistical optics description, one views the radiation field ${\bf E}({\bf r}, t)$ as a stationary (ergodic) random process. As in the case of other stochastic processes, the coherence properties of partially coherent light can be studied by means of appropriate correlation functions. In the X-ray optics setting, one is often interested in the correlation properties in the transverse plane, ${\bf r}(x,y)$, at different locations $s$ along the beamline.  In the theory of partially coherent radiation, an important role is played by the {\it mutual coherence function} $\Gamma({\bf r}_1, {\bf r}_2; \tau)$, a 2-point (also called `2$^{\rm nd}$ order') correlation function introduced as a measure of correlation between the (complex scalar) values of the radiation E-field at two points in the transverse plane, time $\tau$ apart:
\begin{equation}
\Gamma({\bf r}_1, {\bf r}_2; \tau) = \langle E({\bf r}_1, t +\tau) E^*({\bf r}_2, t)    \rangle_T \; .
\end{equation}
The average is over time intervals $T$ long compared to the characteristic time on which the field values fluctuate. 

In practice, it is often convenient to analyze the field correlation properties by using a dual 2-point correlation function, the {\it cross-spectral density} (CSD), defined as the Fourier transform of the mutual coherence function from the time domain to the frequency domain: 
\begin{equation}
\tilde{\Gamma}({\bf r}_1, {\bf r}_2; \omega) = \int_{-\infty}^{\infty} \Gamma({\bf r}_1, {\bf r}_2; \tau) e^{i\omega\tau} d\tau \; .
\end{equation}
The corresponding normalized correlation function, called {\it spectral degree of coherence}, is defined as
\begin{equation}
\mu_s({\bf r}_1, {\bf r}_2; \omega)  \equiv  \frac{\tilde{\Gamma}({\bf r}_1, {\bf r}_2; \omega)}  { \sqrt{\tilde{\Gamma}({\bf r}_1, {\bf r}_1; \omega)} \sqrt{\tilde{\Gamma}({\bf r}_2, {\bf r}_2; \omega)} } \; . 
\end{equation}
(A note on our choice of notation: Many authors use $W({\bf r}_1, {\bf r}_2; \omega)$ to denote CSD; we use $\tilde{\Gamma}({\bf r}_1, {\bf r}_2; \omega)$ instead, and reserve $W$ exclusively for the Wigner functions.  Aside from differences in notation, the definitions in this section are in agreement with those given in~\cite{Vartanyants_2010}. In this paper we restrict discussion to  monochromatic radiation, so the dependence on $\omega$ will be by and large omitted from notation.)

\subsection{Partially Coherent Radiation in the Wigner Function Formalism } 

A comprehensive discussion of the Wigner function formalism for describing partially coherent radiation is given in~\cite{Bazarov}.  In this section we only introduce the definitions and notation used in the present paper. 

In general, synchrotron radiation is partially coherent. The origin of the partial coherence is in the discrete nature of the radiation source, in which a large number of electrons radiate independently, their longitudinal separation being typically large in relation to the radiation wavelength. A single electron will produce a certain radiation pattern transversely. In the case that the electron beam distribution is 
comparable in size to this radiation width, the overall resulting radiation is highly coherent. This is referred to as the ``diffraction limit". Most 3rd generation synchrotron light sources are not diffraction
limited, but the next generation of low-emittance electron rings brings the radiation closer to this fully coherent case.

It is well known that a partially coherent, statistically stationary radiation field can be represented as a sum of coherent modes~\cite{Mandel1995Sep, Vartanyants_2010}. The coherent modes $E_j({\bf r}; \omega)$ are the eigenfunctions of the linear integral operator that has the cross-spectral density as its kernel:
\begin{equation}
\int \tilde{\Gamma}({\bf r}_1, {\bf r}_2; \omega) E_j({\bf r}_1; \omega) d{\bf r}_1 
= \beta_j(\omega) E_j({\bf r}_2; \omega),
\label{eq:coher_modes}
\end{equation}
and the CSD can be written as a sum of contributions from individual coherent modes,
\begin{equation}
\tilde{\Gamma}({\bf r}_1, {\bf r}_2; \omega) = \sum_j \beta_j(\omega) 
E_j^*({\bf r}_1; \omega) E_j({\bf r}_2; \omega). 
\label{eq:CSDfromCMs}
\end{equation}

An alternative, fully equivalent description of the dynamics and correlation properties of the partially coherent radiation can be effected with the Wigner function 
formulation~\cite{Bazarov}. For a fully coherent radiation wavefront, the Wigner function is defined as 
\begin{equation}
W({\bf r}, \boldsymbol\theta) = \frac{1}{\lambda^2} \int^{\infty}_{\infty} E^*({\bf r} -\boldsymbol\xi/2)E({\bf r} +\boldsymbol\xi/2) e^{-i 2 \pi \boldsymbol\theta \boldsymbol\xi/\lambda} d\boldsymbol\xi \; ,
\end{equation}
where $\lambda$ is the radiation wavelength. In the general case of partially coherent radiation, the Wigner function is a sum of contributions from individual coherent modes:
\begin{eqnarray}
&&W({\bf r}, \boldsymbol\theta) = \frac{1}{\lambda^2} \sum_j \beta_j 
\nonumber \\
&& \times \int^{\infty}_{\infty} E_j^*({\bf r} -\boldsymbol\xi/2)
E_j({\bf r} +\boldsymbol\xi/2) e^{-i 2 \pi \boldsymbol\theta 
\boldsymbol\xi/\lambda} d\boldsymbol\xi \; \; .
\end{eqnarray}
In the uncoupled case, one can similarly introduce the Wigner function in $2D$ phase space,
\begin{eqnarray}
&&W(x, \theta) = \frac{1}{\lambda} \sum_j \beta_j \nonumber \\
&&\times \int^{\infty}_{\infty} E_j^*(x -\xi/2)
E_j(x +\xi/2) e^{-i 2 \pi \theta 
\xi/\lambda} d\xi \; \; .
\end{eqnarray} 
In general, a Wigner function describes partially coherent radiation.

In terms of the Wigner function one can define a useful measure of coherence called 
{\it degree of coherence} $\mu$:
\begin{equation}
\mu^2 =  \lambda \int W^2(x, \theta) dx d \theta \; 
\end{equation}
In the fully-coherent case, $\mu = 1$, while in the partially coherent case $\mu < 1$.  

The cross-spectral density and the Wigner function are in a one-to-one relation, related by a pair of integral transforms (the Wigner-Weyl transform pair): 
\begin{equation}
\tilde{\Gamma}({\bf r}_1, {\bf r}_2) 
= \int W\left(\frac{{\bf r}_1 + {\bf r}_2}{2}, \boldsymbol\theta \right) 
e^{i 2 \pi ({\bf r}_1 - {\bf r}_2) \boldsymbol\theta /\lambda} 
d\boldsymbol\theta \; 
\label{eq:WFtoCSD4D}
\end{equation} 
and its inverse in the fully coupled case ($4D$ phase space), and 
\begin{equation}
\tilde{\Gamma}(x_1, x_2) = \int W\left(\frac{x_1 +x_2}{2}, \theta \right) e^{i 2 \pi (x_1 -x_2) \theta /\lambda} d\theta \; 
\label{eq:WFtoCSD2D}
\end{equation}
and its inverse in the uncoupled case ($2D$ phase space). 
Thus, one has the freedom of tracking either the CSD or the Wigner function.  Tracking a Gaussian Wigner function results in a reduced description by means of an envelope model similar to the one used in modeling charged particle beams.

\subsection{Gaussian Wigner Functions}

A Gaussian Wigner function is completely specified by the matrix of its second moments (the covariance matrix $\Sigma$) as well as offsets in position and angle (phase space centroid coordinates), if any. Although not a probability distribution density in the usual sense, a Wigner function can be used in the same manner as a regular phase space distribution density function for the purpose of computing the moments of the distribution~\cite{Bazarov}.  As long as the Wigner function remains Gaussian, it is clearly sufficient to only track its moments up to the second order to have a complete information about it at all times, in contrast to a general Wigner function where the number of moments that have to be tracked is in principle infinite.

We use the following notation for the elements of the $\Sigma$-matrix of the radiation beam: 
\begin{equation} 
\Sigma = 
\begin{pmatrix}
    \sigma_{xx} & \sigma_{x\theta}\\
    \sigma_{x\theta} & \sigma_{\theta\theta}
\end{pmatrix} \; ,
\end{equation} 
where $\sigma_{xx} = \langle x^2 \rangle$, etc., in terms of which the rms emittance is defined as 
\begin{equation} 
\epsilon = \sqrt{\sigma_{xx}\sigma_{\theta\theta} -\sigma_{x\theta}^2} = \left( \det \Sigma \right)^{1/2} \; .
\end{equation} 

In terms of the {\it beam quality factor} $m^2$, which is a measure of coherence related to the degree of coherence $\mu$ as 
\begin{equation}
m^2 \equiv \frac{1}{\mu} \; ,
\end{equation}
the emittance can be written as 
\begin{equation}
\epsilon = m^2 \frac{\lambda}{4\pi} \; .
\label{emittance_lambda_msquared}
\end{equation}
The smallest possible value of the emittance, $\lambda / 4\pi$, obtains only for a fully coherent beam.

For convenience, we also introduce the quantities $\alpha$, $\beta$ and $\gamma$ similar to the Twiss parameters used in beam physics:
\begin{equation} 
\epsilon \beta = \sigma_{xx}, \;\; \epsilon \gamma = \sigma_{\theta\theta}, \;\; \epsilon \alpha = - \sigma_{x\theta}, \; \; \; \; 
\end{equation} 
which are constrained by the identity 
\begin{equation} 
\beta \gamma - \alpha^2 = 1 \; .
\label{twiss_ident} 
\end{equation} 
The inverse of $\Sigma$ is given by
\begin{equation} 
\Sigma^{-1} = \frac{1}{\det \Sigma}
\begin{pmatrix}
    \sigma_{\theta\theta} & -\sigma_{x\theta}\\
   - \sigma_{x\theta} & \sigma_{xx}
\end{pmatrix} 
= \frac{1}{\epsilon}
\begin{pmatrix}
    \gamma & \alpha\\
   \alpha & \beta
\end{pmatrix} \; .
\end{equation} 
The Wigner function for the incoming radiation is then
\begin{equation}
W_i(\vec z) = \frac{1}{2\pi \sqrt{\det\Sigma}} \exp \left(  -\frac{1}{2} \vec{z}^T \Sigma^{-1} \vec{z}  \right)  \; ,
\end{equation}
where in $2D$ phase space $\vec{z} \equiv (x, \theta)^T$, and the normalization convention 
\begin{equation}
\int W(\vec{z}) d\vec{z}  = 1
\end{equation}
is used.

For reference and with a view towards performing the convolution operation in $\theta$ space, we list here three equivalent representations of $W_i(x, \theta)$ in $2D$ phase space:
\begin{equation}
\begin{split}
&W_i(x, \theta) \\
&= \frac{1}{2\pi \sqrt{\det\Sigma}} \exp \left[  -\frac{1}{2\det\Sigma} (\sigma_{\theta\theta} x^2 -2 \sigma_{x\theta}x\theta +\sigma_{xx}\theta^2 )  \right] \\
&= \frac{1}{2\pi \epsilon} \exp \left[  -\frac{1}{2\epsilon} (\gamma x^2 +2 \alpha x\theta +\beta \theta^2 )  \right] \\
&= \frac{1}{2\pi \epsilon} \exp \left[  -\frac{x^2}{2\epsilon\beta}  \right]   \exp \left[  -\frac{\beta}{2\epsilon}  \left( \theta +\frac{\alpha}{\beta}x \right)^2  \right]  \; .
\end{split}
\label{w_i_2D}
\end{equation}

Using the general relation between the radiation Wigner function and cross-spectral density given by Eq.~\eqref{eq:WFtoCSD2D}, we find that for a Gaussian Wigner function $W(x, \theta)$ as in Eq.~\eqref{w_i_2D}, 
\begin{equation}
\begin{split}
\tilde{\Gamma}(x_1, x_2) &\propto e^{-i \frac{\pi \alpha}{\lambda\beta} (x_1 +x_2)(x_1 -x_2)} \\ 
&\times \exp \left(  -\frac{(x_1 +x_2)^2}{8\epsilon\beta} -\frac{(x_1 -x_2)^2}{2 (\beta\lambda^2/4\pi^2\epsilon)}  \right) \; ,
\end{split}
\label{eq:Gamma_GSM}
\end{equation}
whence 
\begin{equation}
\begin{split}
\mu_s(x_1, x_2) =&  \exp\left[ -i \frac{\pi \alpha}{\lambda\beta} (x_1 +x_2)(x_1 -x_2) \right. \\
&\left. -\frac{(x_1 -x_2)^2}{2} \left( \frac{-1}{4\sigma_{xx}} +\frac{4\pi^2}{\lambda^2}\frac{\epsilon^2}{\sigma_{xx}}   \right)   \right] \\
&\propto \exp\left[ -\frac{(x_1 -x_2)^2}{2\xi_x^2} \right]\; .
\end{split}
\label{eq:mu_s_gauss}
\end{equation}
That this is the same result as arises in the Gaussian Schell model~\cite{Vartanyants_2010,Bazarov,Mandel1995Sep}, is no coincidence.  Indeed, Gaussian Wigner function implies Gaussian Schell model, and {\it vice versa}. \footnote{See reference \cite{Simon1985Apr} for a general treatment of the relation of 4D Gaussian Wigner functions and Gaussian Schell models.}

Equation~\eqref{eq:mu_s_gauss} serves to define (for a Gaussian Wigner function) the {\it transverse coherence length} $\xi_x$, which in terms of the rms beam size $\sigma_x = \sqrt{\sigma_{xx}}$ is given by
\begin{equation}
\xi_x = \frac{2\sigma_x}{\sqrt{ \frac{\epsilon^2}{(\lambda /4\pi)^2}  -1}}
= \frac{2\sigma_x}{\sqrt{m^4 -1}} \; .
\label{eq:coher_len_x}
\end{equation}
Note that this expression for coherence length is consistent with the result given by Onuki and Elleaume\cite{Hideo2002Oct} Eqn. [142].

In the non-separable case, and allowing for offsets $\vec{z}_0$ of the beam centroid, the expressions for a Gaussian Wigner function in terms of the $4 \times 4$ correlation matrix $\Sigma$ are quite similar, 

\begin{equation}
W_i(\vec{r}, \vec{\theta}) = \frac{1}{2\pi \sqrt{\det\Sigma}} \exp \left(  -\frac{1}{2} (\vec{z} -\vec{z}_0)^T \Sigma^{-1} (\vec{z} -\vec{z}_0) \right)  \; ,
\end{equation}
except that now that the phase space is $4D$, one has to specify the ordering of the phase space variables.  In this paper, we use the ordering convention such that 
\begin{equation}
\vec{z} = 
\begin{pmatrix}
    \vec{r} \\
    \vec{\theta} 
\end{pmatrix} \; ,
\end{equation} 
where 
\begin{equation}
\vec{r} = 
\begin{pmatrix}
    x \\
    y  
\end{pmatrix} \; 
\end{equation} 
and 
\begin{equation}
\vec{\theta} = 
\begin{pmatrix}
    \theta_x \\
    \theta_y 
\end{pmatrix} \; .
\end{equation} 
In what follows, we write the inverse of the correlation matrix $\Sigma$ in terms of four $2\times2$ auxiliary matrices $A$, $B$, $C$, and $D$ such that 
\begin{equation} 
\Sigma^{-1} = 
\begin{pmatrix}
    A & B\\
    C & D
\end{pmatrix} \; .
\end{equation} 
Because $\Sigma$ is symmetric, we have $\Sigma^{-1} = (\Sigma^{-1})^T$, and therefore $A = A^T$, $D = D^T$, and $B = C^T$. It follows that 
\begin{equation}
-\frac{1}{2} {\vec{z}}^T \Sigma^{-1} \vec{z} =  -\frac{1}{2} \left(  {\vec{\theta}}^T D \vec{\theta}  +2(C \vec{r})^T \vec{\theta}  +{\vec{r}}^T A \vec{r} \right)  \; 
\end{equation}
for the case of no offset in any of the phase space coordinates, with an obvious generalization for the case of non-zero centroid offsets.

\section{Passage through a Hard-Edge Aperture}
\label{sec:03}
After passing through a physical (hard-edge) aperture, a Gaussian radiation Wigner function becomes non-Gaussian.  In addition, there is a loss of radiation flux which is, for the case of a hard-edge aperture, a purely geometric effect, the outer part of the beam being blocked by the aperture diaphragm.  It is possible to calculate analytically the effects of a hard-edge aperture on an incoming Gaussian Wigner function; we present the results below.

\subsection{Loss of Flux} 

The flux loss for an ideal straight-hard-edge aperture extending in $x$ from $-a_h$ to $a_h$ is found by purely geometric considerations. Let us consider the separable case. If the intensity distribution $I_i$ immediately before the aperture is given by
\begin{equation}
I_i = I_0 \frac{1}{\sqrt{2\pi}\sigma_x}\exp \left( -\frac{(x-\mu)^2}{2\sigma_x^2} \right) \; ,
\end{equation}
the ratio of the flux $F_f$ immediately after the aperture to the flux $F_i$ before the aperture will be
\begin{equation}
\begin{split} 
F_f / F_i =  \frac{1}{\sqrt{2\pi}\sigma_x} \int^{a_h}_{-a_h} \exp \left( -\frac{(x-\mu)^2}{2\sigma_x^2} \right) dx \\
= \frac{1}{2} \left[ \erf\left(\frac{a_h +\mu}{\sqrt{2}\sigma_x}\right)  +\erf\left(\frac{a_h -\mu}{\sqrt{2}\sigma_x}\right)   \right] \; .
\end{split} 
\end{equation}
When the incident beam is centered on the aperture ($\mu = 0$), this simplifies to 
\begin{equation}
F_f / F_i = \erf\left(\frac{a_h}{\sqrt{2}\sigma_x}\right) \; .
\end{equation}
The corresponding result for a rectangular aperture in two spatial dimensions is the product of the 1D results in $x$ and $y$.

\subsection{Aperture ``Wigner Function''}

In addition to the loss of flux, the loss of the Gaussian character of the radiation Wigner function occurs in traversing a physical hard-edge aperture.  Quantitative description of the effect necessitates the introduction of the aperture ``Wigner function''. We put the term in quotes, since this formally introduced function differs from the ``true'' Wigner function of the radiation beam not just in its physical content/meaning but also in some purely mathematical aspects; in particular, the aperture ``Wigner function'' is not normalized to unity. 
 
Let us consider the separable case ($2D$ phase space). The starting point is to introduce the aperture transfer function $t(x;a)$ for the E-field of the radiation wavefront, in terms of which the the radiation field $\vec{E}_i(x)$ before the aperture is related to the radiation field $\vec{E}_f(x)$ after the aperture as 
\begin{equation}
\vec{E}_f(x) = t(x; a)\vec{E}_i(x) \, .
\end{equation}
Here, $a$ stands for parameters that specify the aperture.  
In the region of the transverse plane not obstructed by the physical aperture, $t(x; a) = 1$.

The aperture ``Wigner function'' is formally defined from its transfer function in the same way as the radiation Wigner function is defined from the electric field of a coherent monochromatic wavefront:
\begin{equation}
W_a(x, \theta) = \frac{1}{\lambda} \int^{\infty}_{\infty} t^*(x -\xi/2; a)t(x +\xi/2; a) e^{-i 2 \pi \theta \xi/\lambda} d\xi .
\label{eq:ap_WF_general}
\end{equation}
As mentioned, unlike the Wigner function for the radiation field, the aperture ``Wigner function'' is not normalized to unity. 

For a thin-screen, straight-edge aperture that extends in $x$ from $-a_h$ to $a_h$, one finds that 
\begin{equation}
W_a(x, \theta) = 
\left\{ \begin{aligned} 
   &\frac{\sin\left( \frac{4(a_h -|x|)}{\lambda} \pi \theta   \right)}{\pi \theta},  \;\; -a_h \le x \le a_h  \; ,\\
  &0 \; , |x| > a_h \; .
\end{aligned} \right.
\label{eq:WFap2Dhardedge}
\end{equation}
(Note the presence of dependence on the radiation wavelength $\lambda$.)

\subsection{General Results in 2D Phase Space} 

The (non-Gaussian) Wigner function after a thin-screen hard-edge aperture can be calculated as a convolution in $\vec{\theta}$ 
of the incident Gaussian Wigner function of the radiation with the ``Wigner function'' of the aperture.  In the separable case, allowing for non-zero Twiss $\alpha$ and misalignments $x_0$ and $\theta_0$ in position and angle, the radiation Wigner function $W_i(x, \theta)$ before the aperture is given by a slight generalization of Eq.~\eqref{w_i_2D}, 
\begin{equation}
\begin{split}
&W_i(x, \theta) =\frac{1}{2\pi\epsilon} \exp \left(  -\frac{(x-x_0)^2}{2\epsilon\beta}  \right) \\
&\times    \exp \left(  -\frac{1}{2(\epsilon/\beta)}  \left( \theta -\theta_0+\frac{\alpha}{\beta}(x -x_0)  \right)^2  \right) \; ,
\end{split}
\label{eq:W_i_2D_general} 
\end{equation}
and the ``Wigner function" for the hard-edge aperture, by Eq.~\eqref{eq:WFap2Dhardedge}.
We make use of the convolution theorem, \textit{viz},
\begin{equation}
f \ast g = IFT[FT[f] \cdot FT[g]] \; ,
\end{equation}
to compute the convolution in question,
\begin{equation}
W_f(x, \theta) = W_i(x, \theta) *_{\theta} W_a(x, \theta)  \; ,
\end{equation}
where $*_{\theta}$ denotes the convolution operation in $\theta$ alone.
The definitions of the Fourier transform and the inverse Fourier transform used in this paper are, respectively,
\begin{equation}
\tilde{f}(t) \equiv FT[f(\theta)] = \int f(\theta) e^{-i 2\pi t \theta} d\theta 
\label{eq:ft}
\end{equation}
and 
\begin{equation}
IFT[\tilde{f}(t)] = \int \tilde{f}(t) e^{i 2\pi \theta t} dt  \; .
\label{eq:ift}
\end{equation}

Details of the calculation are given in Appendix~\ref{app:A}. The resulting expression for the radiation Wigner function immediately after the hard-edge aperture in 2D phase space can be written as 
\begin{equation}
\begin{split}
W_f(x, \theta) =& W_i(x, \theta)\ \Pi_{a_h}(x) \\
& \times \operatorname{Re} \left\{ \erf \left[ \sqrt{\frac{2\epsilon}{\beta}} \left( \frac{2\pi (a_h -|x|)}{\lambda}   \right. \right. \right. \\
& \left. \left. \left. +i \left( \frac{\beta}{2\epsilon}(\theta -\theta_0) +\frac{\alpha}{2\epsilon}(x-x_0)   \right)   \right) \right] \right\}  \; ,
\end{split}
\label{eq:W_f_hardedge_text} 
\end{equation}
where 
\begin{equation}
\Pi_a(x) = 
\left\{ \begin{aligned} 
   &1,  \;\; |x| < a  \; ,\\
  & 1/2 \; , \; |x| = a > 0 \; , \\
  &0 \; , |x| > a \; .
\end{aligned} \right.
\end{equation}
It is easily seen that with $\alpha = 0$ and zero offsets $x_0$ and $\theta_0$, this result reduces to the expression previously reported in~\cite{Nash2014_PRAB.24.010702}. 

This result is for the Wigner function immediately after the aperture. The Wigner function in a drift at a distance $L_d$ past the aperture is given by $W_f(x -\theta L_d, \theta)$ (see Eq.~\eqref{eq:W_f_at_Ld_hardedge} below).  For a particular choice of the system parameters, the evolution of the Wigner function in a drift following the hard-edge aperture is illustrated in Fig.~\ref{fig:WF_10_30_cm_and_proj_on_x:a} and Fig.~\ref{fig:WF_10_30_cm_and_proj_on_x:b}.

\section{Gaussian Aperture}
\label{sec:04}
As already mentioned, the radiation Wigner function after an aperture is computed by performing a convolution, in the angle variable $\vec{\theta}$ only, of the Wigner function of the incident wave front with the formally defined ``Wigner function'' of the aperture. 
In this section we consider the passage through a Gaussian aperture, in which case the Gaussian nature of the Wigner function is preserved.  We treat the separable, $2D$ phase space case first, since it provides a clearer picture of the relevant physics, in addition to often being the model of choice in practice.  We then present the general result for the coupled, $4D$ phase space with the beam centroid offset in both position and angle, for a non-matched beam (in the sense that the Twiss $\alpha$ is non-zero). 

\subsection{Gaussian Aperture ``Wigner Function''} 

We define the Gaussian aperture as one having the transfer function 
\begin{equation}
t(x; a_g) = \exp(-x^2/2a_g^2) \; , 
\end{equation}
the pre-factor being set to unity so that the limit of an infinitely large $a_g$ corresponds to free-space propagation. 
The ``Wigner function'' for a Gaussian aperture is then given by Eq.~\eqref{eq:ap_WF_general} with the above $t(x; a_g)$, {\it i.e.}, 
\begin{equation}
W_a(x, \theta) =  \frac{2\sqrt{\pi} a_g}{\lambda} \exp \left( -\frac{x^2}{a_g^2} \right) \exp \left( -\frac{\theta^2}{(\lambda/2\pi a_g)^2} \right) \; , 
\label{eq:apertW} 
\end{equation}
where $\lambda$ is the radiation wavelength. Thus, aside from $\lambda$, the Gaussian aperture is defined by a single parameter $a_g$. Note that the formally defined rms emittance in this case is 
\begin{equation}
\epsilon = \sqrt{ \langle x^2 \rangle \langle \theta^2 \rangle } = \frac{\lambda}{4\pi} \; , 
\end{equation}
consistent with the fact that the Gaussian E-field distribution is tantamount to fully coherent radiation. 

\subsection{The Case of 2D Phase Space with No Centroid Misalignments} 

A straightforward calculation, the details of which are given in Appendix~\ref{app:B}, yields the elements of the $\Sigma_f$ matrix after a Gaussian aperture in terms of the effective size parameter of the aperture $a_g$ and the elements of the $\Sigma$ matrix before the aperture: 
\begin{equation}
\sigma_{xx,f} 
=  \frac{a_g^2/2}{\sigma_{xx} +a_g^2/2} \sigma_{xx} \; , 
\label{eq:sigma_xx_f}
\end{equation} 
\begin{equation}
\sigma_{x\theta,f} 
=  \frac{a_g^2/2}{\sigma_{xx} +a_g^2/2} \sigma_{x\theta} \; , 
\label{eq:sigma_xt_f}
\end{equation} 
\begin{equation}
\sigma_{\theta\theta,f} 
= \sigma_{\theta\theta} -\frac{\sigma_{x\theta}^2}{\sigma_{xx} +a_g^2/2} 
+\left( \frac{\lambda}{4\pi} \right)^2 \frac{1}{a_g^2/2} \; .
\label{eq:sigma_tt_f}
\end{equation}

We note in passing that 
\begin{equation}
\frac{\sigma_{x\theta}}{\sigma_{xx}} = const 
\end{equation}
in going through the aperture, for all values of $a_g$.  It is easily verified that, as expected, in the limit $a_g \to \infty$ (corresponding to free-space propagation) the elements of the covariance matrix and the emittance do not change. It is instructive to re-write Eq.~\eqref{eq:eps2_f} that relates the values of emittance before and after the aperture in the form 
\begin{equation}
\epsilon_f^2 -\left( \frac{\lambda}{4\pi} \right)^2 = \left[  \epsilon^2 -\left( \frac{\lambda}{4\pi} \right)^2  \right] \frac{a_g^2/2}{\sigma_{xx} +a_g^2/2} \; ,
\label{emittance_change_aperture}
\end{equation}
whence it is also clear that 
\begin{equation}
\lim_{a_g \to 0} \epsilon_f = \frac{\lambda}{4\pi} \; .
\end{equation}
This means the radiation becomes fully coherent in the limit of infinitely small aperture size.  (Of course, the trade-off is that the transmitted flux is approaching zero in this limit.)

\subsection{Transverse coherence length does not change in passing through Gaussian aperture}
\label{coherence_length_through_aperture}

Using Eq.~\eqref{eq:coher_len_x} for the transverse coherence length $\xi_x$ together with Eqs.~\eqref{eq:sigma_xx_f}  and \eqref{emittance_change_aperture} that relate $\sigma_{xx}$ and emittance before and after a Gaussian aperture, we find that 
\begin{equation}
\begin{split}
\frac{1}{\xi_f^2} = \frac{1}{4\sigma_{xx,f}} \left[ \frac{\epsilon_f^2}{(\lambda/4\pi)^2} -1   \right]  \\
= \frac{1}{4\sigma_{xx}} \left[ \frac{\epsilon^2}{(\lambda/4\pi)^2} -1   \right] = \frac{1}{\xi^2} \; ,
\end{split}
\label{eq:invar_xi}
\end{equation}
{\it i.e.}, $\xi_x$ remains constant in traversing the aperture.  This result may seem somewhat counter-intuitive. However, the rms transverse beam size $\sigma_x$ becomes smaller in passing through the aperture, and so the ratio $\xi / \sigma_x$ becomes larger. By this measure, the beam does become more coherent after the aperture.

It follows from Eq.~\eqref{eq:coher_len_x} that in free space following the aperture, the transverse coherence length grows in proportion to the rms beam size, 
\begin{equation}
\frac{\xi_x(s)}{\sigma_x(s)} = const.
\end{equation}

\subsection{The Case of 2D Phase Space with Misalignments in Angle and Position}

A straightforward calculation, the details of which are presented in Appendix~\ref{app:B_w_offsets}, yields a relation between the elements of the covariance matrix and the offsets in angle and position before and after a Gaussian aperture. The result is that the covariance matrices before and after the aperture are related in the same way as in the case of no misalignments, Eqs.~\eqref{eq:sigma_xx_f} - \eqref{eq:sigma_tt_f}, while the expressions for offsets after the aperture are given by  
\begin{equation}
x_{0,f} = \left( 1 -\frac{\sigma_{xx,f}}{a_g^2/2} \right) x_0 
= \frac{\sigma_{xx,f}}{\sigma_{xx}} x_0
\label{eq:x_offset_f}
\end{equation}
and
\begin{equation}
\theta_{0,f} = \theta_0 -\frac{\sigma_{x\theta, f}}{a_g^2/2} x_0 \; .
\end{equation}

\subsection{The Case of 4D Phase Space with No Misalignments}

When the horizontal and vertical dimensions are coupled and the aperture's ``Wigner function'' is known, a convolution-in-$\boldsymbol\theta$ procedure that parallels the one used for the 2D phase space is used to calculate analytically the radiation Wigner function after the aperture.  For a Gaussian aperture, the convolution with the Gaussian Wigner function of the incident wavefront results in a radiation Wigner function that is again Gaussian.  Writing the \textit{inverse} of the covariance matrix $\Sigma^{-1}$ at entrance to the aperture as an arrangement of four $2 \times 2$ blocks $A$, $B$, $C$, and $D$, 
\begin{equation} 
\Sigma^{-1} = 
\begin{pmatrix}
    A & B\\
    C & D
\end{pmatrix} \; ,
\end{equation} 
and making use of the symmetry properties of $\Sigma$, we find the corresponding $2 \times 2$ blocks $A_f$, $B_f$, $C_f$, and $D_f$ of the inverse covariance matrix after the aperture:
\begin{equation}
A_f = A +A_A -BD^{-1}C +BD^{-1} (D^{-1} +D_A^{-1})^{-1} D^{-1}C   \; ,
\label{eq:Af_text} 
\end{equation}
\begin{equation}
B_f = BD^{-1} (D^{-1} +D_A^{-1})^{-1}  \; ,
\end{equation} 
\begin{equation}
C_f = (D^{-1} +D_A^{-1})^{-1} D^{-1}C = B_f^T  \; ,
\end{equation}
and 
\begin{equation}
D_f = (D^{-1} +D_A^{-1})^{-1}  \; ,
\label{eq:Df_text} 
\end{equation}
where, in terms of the parameters $a_x$ and $a_y$ that specify the Gaussian aperture in the $x$ and $y$ directions, 
\begin{equation}
A_A = 
 \begin{pmatrix}
    2/a_x^2 & 0\\
    0 & 2/a_y^2
\end{pmatrix} \; 
\label{eq:AA_text} 
\end{equation} 
and 
\begin{equation}
D_A^{-1} = \frac{\lambda^2}{8\pi^2} 
 \begin{pmatrix}
    1/a_x^2 & 0\\
    0 & 1/a_y^2
\end{pmatrix} \; 
= \left(\frac{\lambda}{4\pi}\right)^2  A_A.
\label{eq:invDA_text} 
\end{equation}
The details of the calculation can be found in Appendix~\ref{app:C}. In practice, the covariance matrix $\Sigma_f$ after the aperture is then computed via a numerical inversion procedure.

These formulae are being implemented to include apertures in the beam statistics report produced by the Sirepo web interface for the SHADOW code \cite{SirepoShadow, SanchezdelRio_2011}.

\subsection{The Case of 4D Phase Space with Misalignments in Angle and Position} 

It is also possible to derive analytically the result for a Gaussian Wigner function traversing a Gaussian aperture in the general coupled $4D$ phase space case, that is, when there are misalignments in angle and position in both vertical and horizontal dimensions. The details of the derivation are in Appendix~\ref{app:D}.  The result is that the elements of the $\Sigma$-matrix after the aperture are related to those before the aperture in exactly the same way as in the case of no misalignments, {\it i.e.}, by Eqs.~\eqref{eq:Af_text}-\eqref{eq:Df_text} with the auxiliary relations given by Eqs.~\eqref{eq:AA_text}and \eqref{eq:invDA_text}.  For the vector of misalignments $\vec{z}_{0f}$ immediately after the aperture one finds a simple expression in terms of the offsets $\vec{z}_0$ before the aperture and the elements of the correlation matrix $\Sigma_f$ {\it after} the aperture, 
\begin{equation}
\vec{z}_{0f} = \vec{z}_0 -\Sigma_f 
\begin{pmatrix}
    A_A\vec{r}_0 \\
    \vec{0} 
\end{pmatrix} \; .
\end{equation}

\begin{figure*}[!htb] 
  \begin{subfigure}[b]{0.5\linewidth}
    \centering 
    \includegraphics[width=0.95\linewidth]{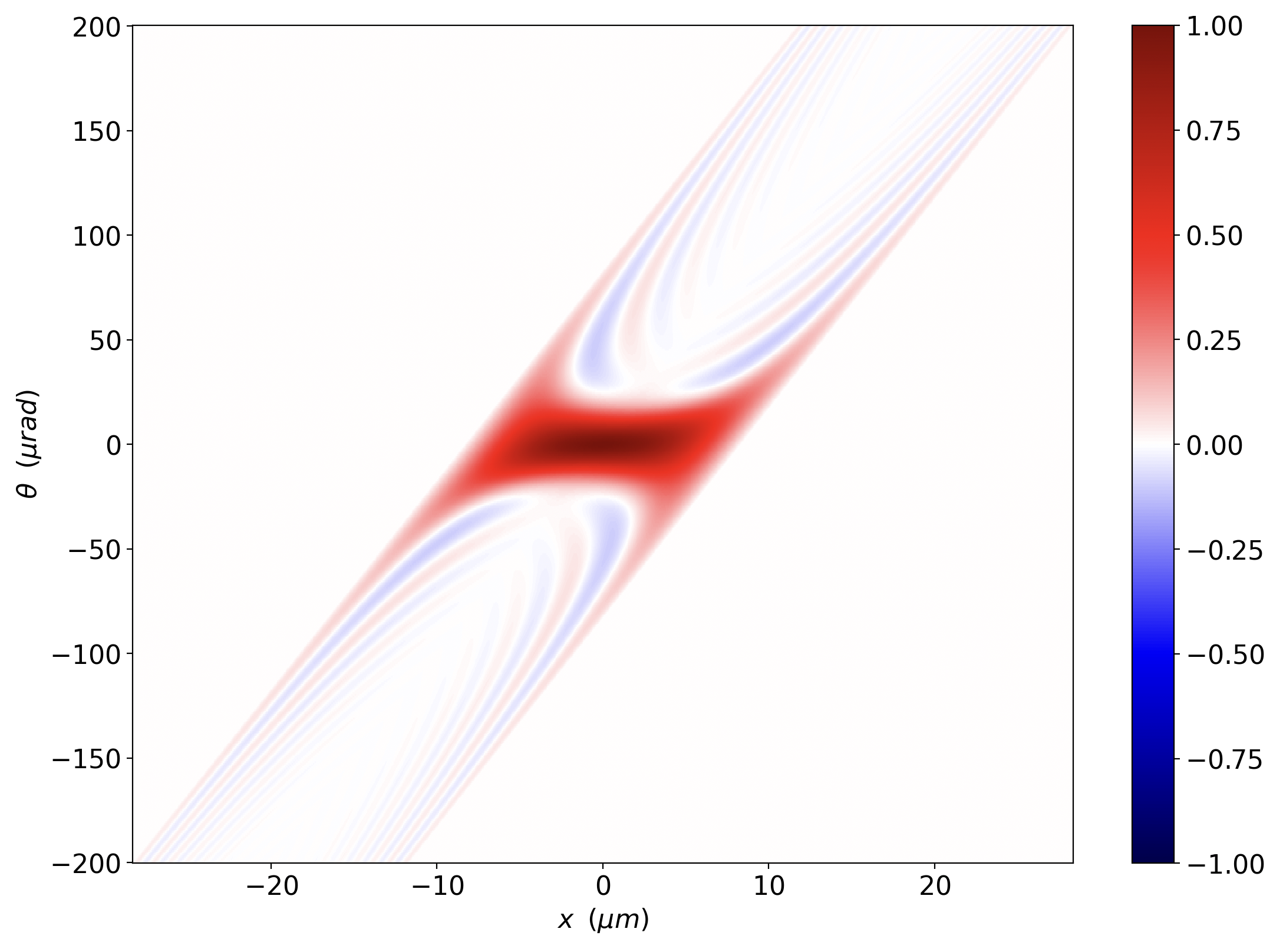} 
    \caption{$W_f(x, \theta)/W_f(0,0)$, 10 cm from the aperture} 
    \label{fig:WF_10_30_cm_and_proj_on_x:a} 
    \vspace{2ex}
  \end{subfigure}
  \begin{subfigure}[b]{0.5\linewidth}
    \centering 
    \includegraphics[width=0.95\linewidth]{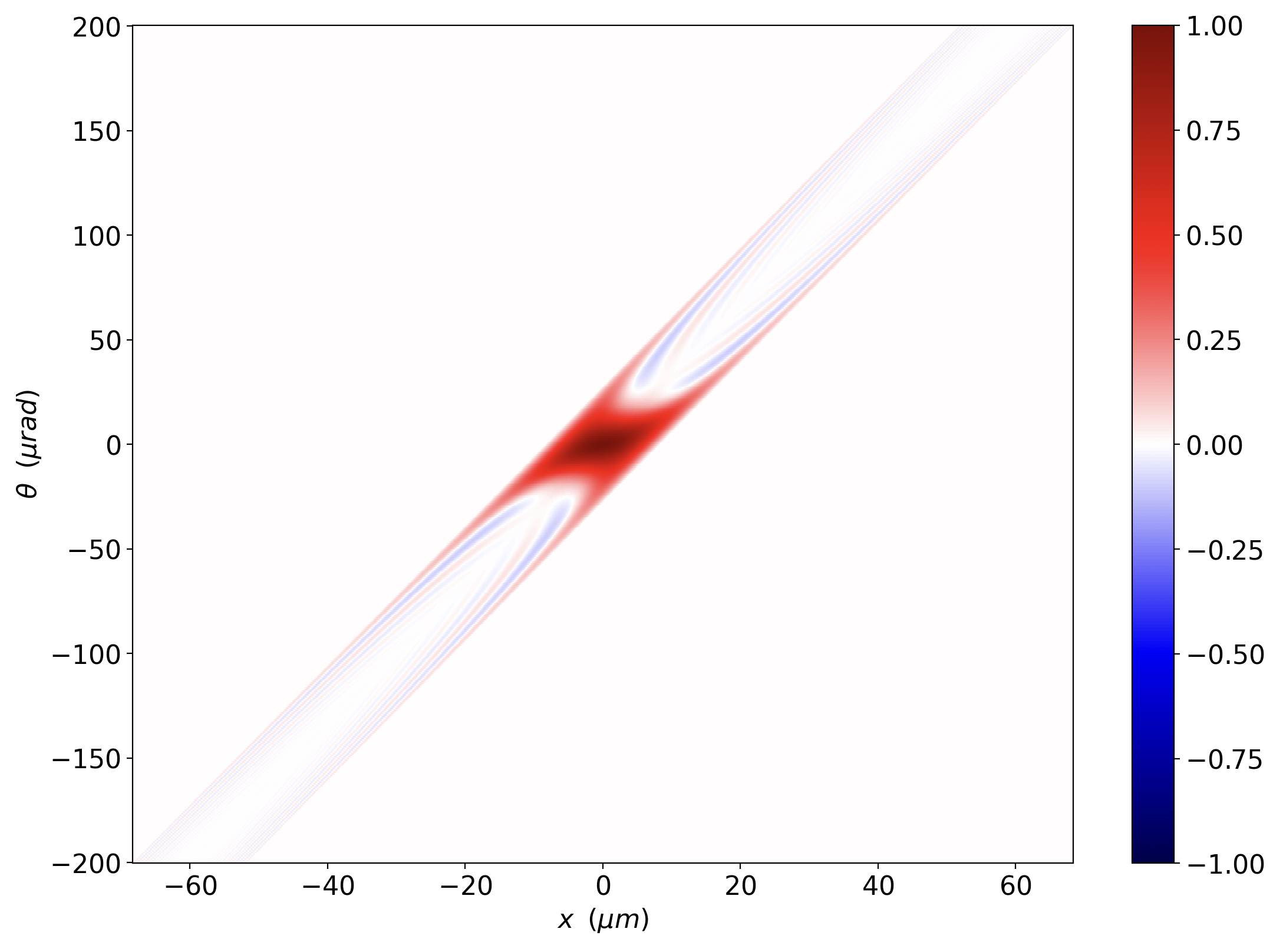} 
    \caption{$W_f(x, \theta)/W_f(0,0)$, 30 cm from the aperture} 
    \label{fig:WF_10_30_cm_and_proj_on_x:b} 
    \vspace{2ex}
  \end{subfigure} 
  \begin{subfigure}[b]{0.5\linewidth}
    \centering 
    \includegraphics[width=0.95\linewidth]{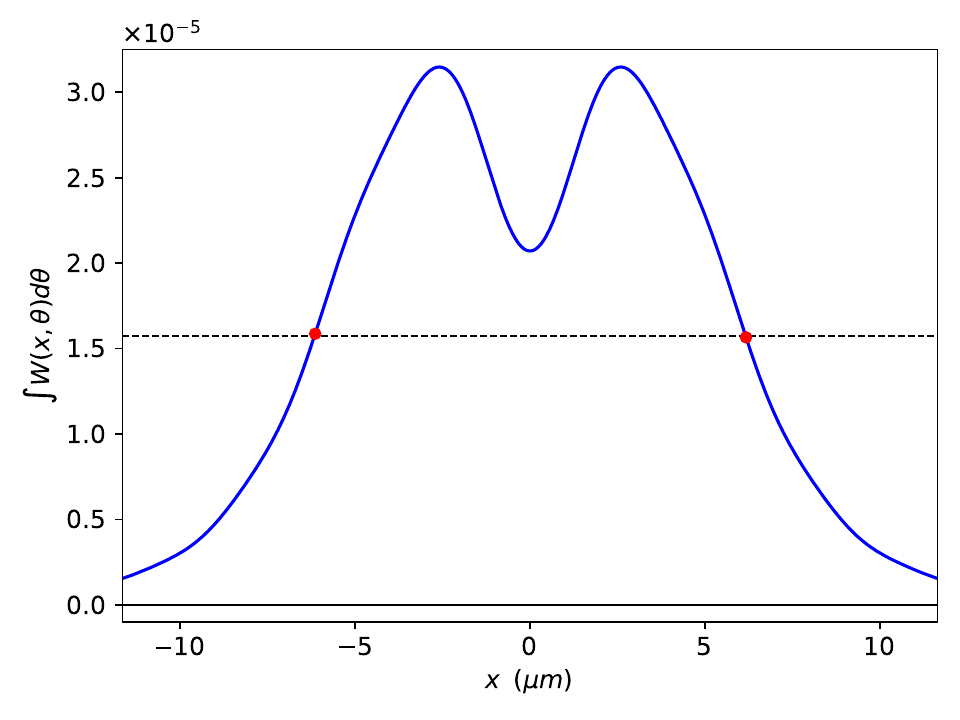} 
    \caption{Projection of the Wigner function on the $x$ axis, 10 cm from the aperture (arbitrary units)} 
    \label{fig:WF_10_30_cm_and_proj_on_x:c} 
  \end{subfigure}
  \begin{subfigure}[b]{0.5\linewidth}
    \centering 
    \includegraphics[width=0.95\linewidth]{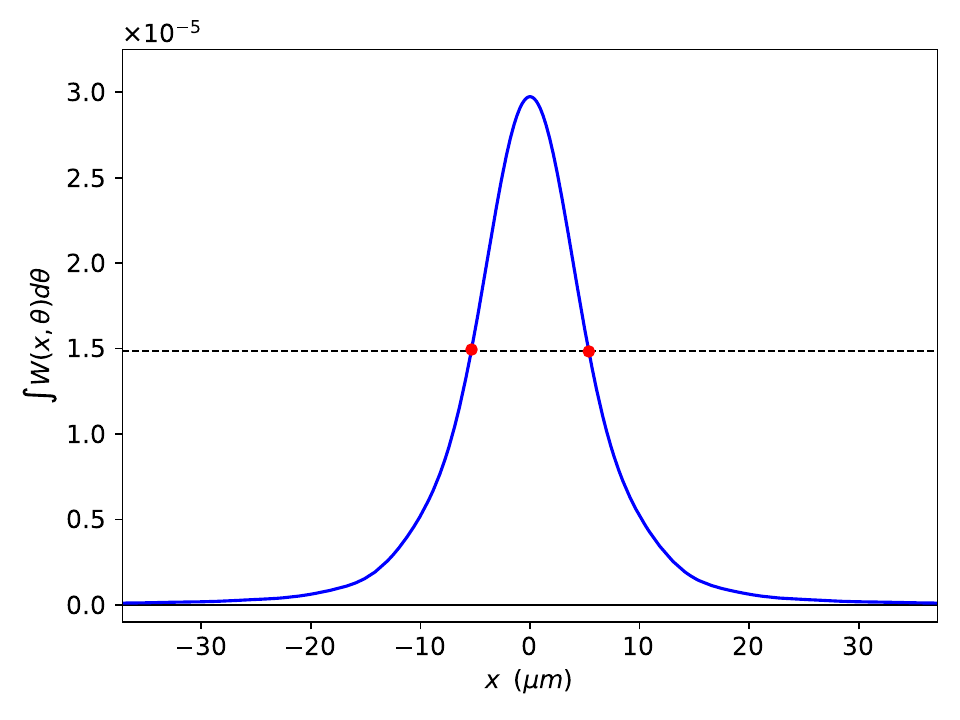} 
    \caption{Projection of the Wigner function on the $x$ axis, 30 cm from the aperture (arbitrary units)} 
    \label{fig:WF_10_30_cm_and_proj_on_x:d} 
  \end{subfigure} 
  \caption{Evolution of the partially coherent radiation Wigner function after passing through a hard-edge aperture (see text). (Same vertical but different horizontal scale in panels (a) and (b); horizontal scale in panels (c) and (d) is chosen for improved visibility and differs from the horizontal scale in the corresponding panels (a) and (b). Dashed line in panels (c) and (d) marks the half-maximum level.)}
  \label{fig:WF_10_30_cm_and_proj_on_x} 
\end{figure*}

\section{Parameters of the Gaussian Approximation}
\label{sec:05}

After traversing a physical hard-edge aperture, the Wigner function of the radiation wavefront is no longer Gaussian.  However, to the extent that our interest is in the parameter regime where modeling accuracy up to and including the second moments of the Wigner function is sufficient, we can construct a Gaussian approximation by properly relating its parameters to the parameters of the physical, non-Gaussian Wigner function. 
As we describe below, the matching of parameters can be done in the far-field regime. In practice, as we illustrate with a numerical example in this section, the distance of transitioning into the far-field regime is often very much smaller than the distance from the aperture to the next optical element, a circumstance that favors our matching-in-the-far-field approach to finding the parameters of the Gaussian approximation. 

Because the momentum-like variable $\theta = const$ in free space, the Wigner function past the aperture, a distance $L_d$ into a drift, is related to the Wigner function immediately after the aperture as 
\begin{equation}
W_f(x, \theta; L_d) = W_f(x -\theta L_d, \theta; L_d = 0) \; .
\label{eq:W_f_at_Ld_general}
\end{equation}
Using Eq.~\eqref{eq:W_f_hardedge_text} for the Wigner function immediately after a hard-edge aperture, we find 
\begin{equation}
\begin{split}
&W_f(x, \theta; L_d) = W_i(x -\theta L_d, \theta)\ \Pi_{a_h}(x -\theta L_d) \\
& \times \operatorname{Re} \left\{ \erf \left[ \sqrt{\frac{2\epsilon}{\beta}} \left( \frac{2\pi (a_h -|x -\theta L_d|)}{\lambda}   \right. \right. \right. \\
& \left. \left. \left. +i \left( \frac{\beta}{2\epsilon}(\theta -\theta_0) +\frac{\alpha}{2\epsilon}(x-x_0 -\theta L_d)   \right)   \right) \right] \right\}  \; ,
\end{split}
\label{eq:W_f_at_Ld_hardedge}
\end{equation} 
with $W_i(x, \theta)$ the Gaussian Wigner function immediately before the aperture, Eq.~\eqref{eq:W_i_2D_general}.

\begin{figure}[!htb]
   \centering
   \includegraphics*[width=1.0\columnwidth]{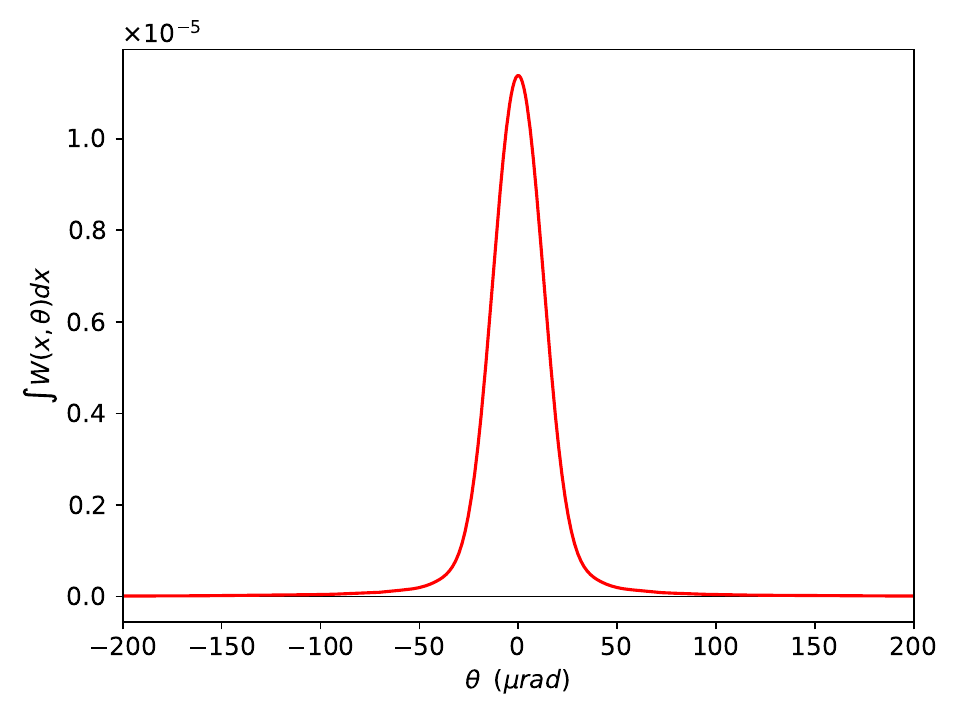}
   \caption{Projection of the Wigner function after a hard-edge aperture on the $\theta$ axis (same for all distances from the aperture).}
   \label{fig:proj_on_theta}
\end{figure}

\begin{figure*}[!htb] 
  \begin{subfigure}[b]{0.5\linewidth}
    \centering
    \includegraphics[width=0.95\linewidth]{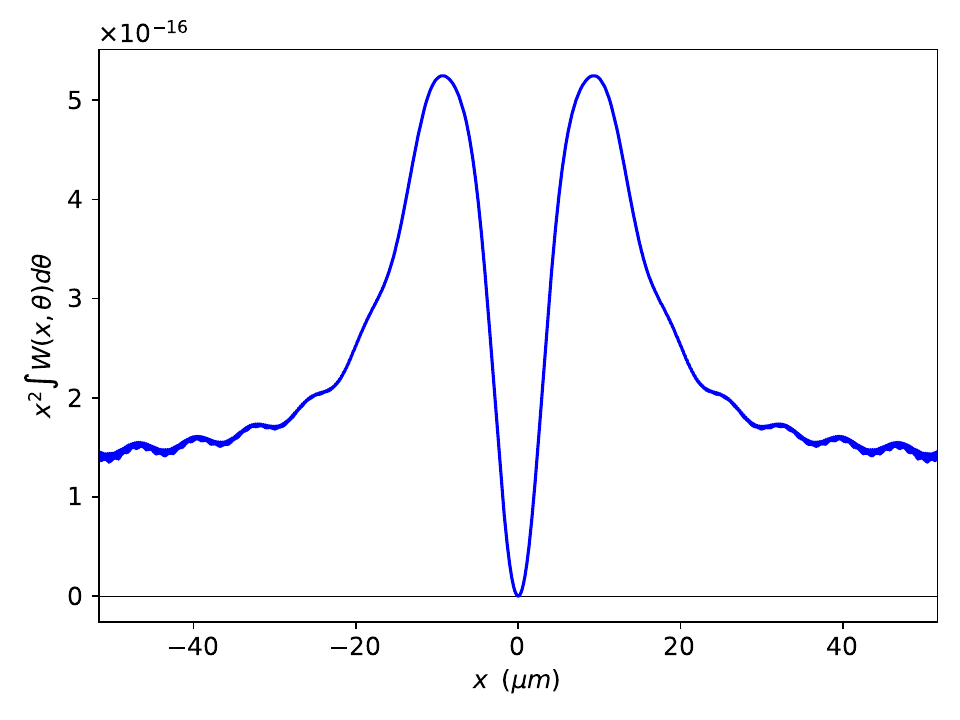}
    \caption{$x^2 \int W_f(x, \theta)d\theta$, 30 cm from the aperture}
    \label{fig:lorentzian:a} 
  \end{subfigure}
  \begin{subfigure}[b]{0.5\linewidth}
    \centering 
    \includegraphics[width=0.95\linewidth]{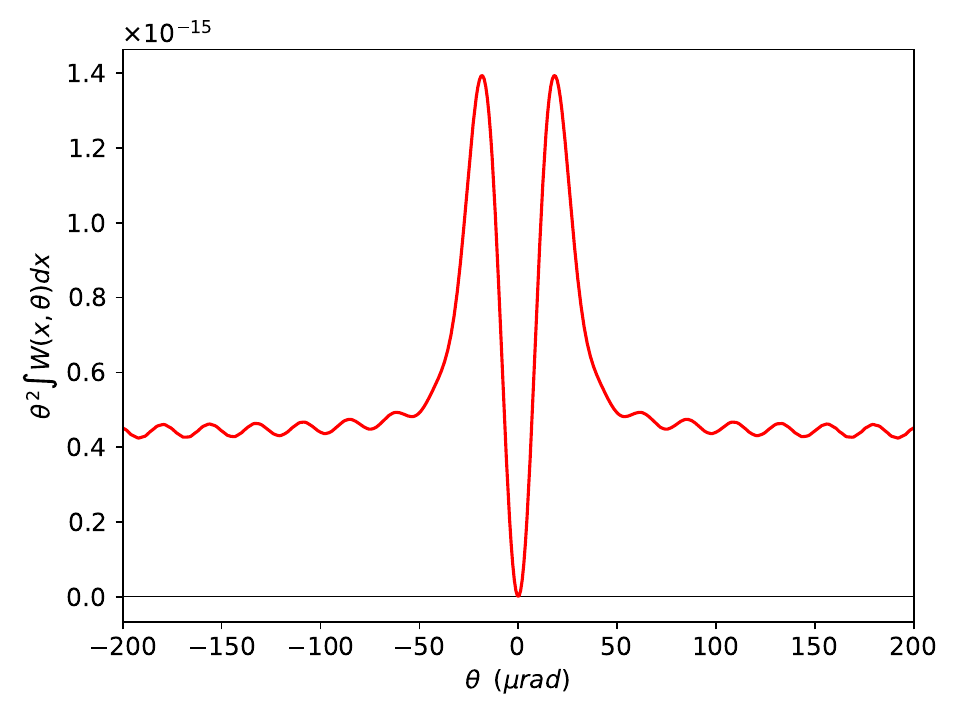} 
    \caption{$\theta^2 \int W_f(x, \theta)dx$, at all distances from the aperture}
    \label{fig:lorentzian:b} 
  \end{subfigure} 
  \caption{Projections on the $x$ and $\theta$ axes of the Wigner function after a hard-edge aperture exhibit an approximately Lorentzian behaviour for large values of the argument, with infinite formally-defined second moments.}
  \label{fig:lorentzian} 
\end{figure*}

Figures \ref{fig:WF_10_30_cm_and_proj_on_x:a} and \ref{fig:WF_10_30_cm_and_proj_on_x:b} show the Wigner function $10$ cm and $30$ cm from the aperture, for a test case where $m^2 = 5$ (recall Eq.~\eqref{emittance_lambda_msquared}), the aperture width is $2 a_h = 16.7\mu$m, $\lambda = 3.98$ $10^{-10}$ m, and $\sigma_{x \theta} = 0$ at the aperture, with zero misalignments in position and angle.  It follows from Eq.~\eqref{eq:W_f_at_Ld_general} that the evolution of the Wigner function in a drift is a shearing parallel to the $x$ axis, so that the projection on the $\theta$ axis, shown in Fig.~\ref{fig:proj_on_theta}, remains unchanged.  The evolution of the intensity (projection of the Wigner function onto the $x$ axis) is somewhat more interesting, as illustrated by the snapshots at $10$ cm and $30$ cm from the aperture in Figures~\ref{fig:WF_10_30_cm_and_proj_on_x:c} and \ref{fig:WF_10_30_cm_and_proj_on_x:d}.  The initially double-peaked structure evolves into a single-peak distribution as the transition to the far-field regime occurs.  Once in the far field, the qualitative similarity with the shapes of the projections on the coordinate axes of the Gaussian Wigner function suggests the possibility of matching the second moments and the centroid coordinates as a way of relating the parameters of the Gaussian approximation to those of the physical Wigner function. However, a closer examination reveals that the projections of the exact Wigner function on both the $x$ and $\theta$ axes in the far field are approximately Lorentzian for large values of $|x|$ and $|\theta|$, as illustrated in Figure~\ref{fig:lorentzian}, so that the formally defined second moments of the physical Wigner function are infinite.  At the same time, the full width at half maximum (FWHM) parameter is well defined for projections of both the exact Wigner function and the Gaussian approximation, and appears to provide a sufficient quantitative measure of the Wigner function footprint in the setting for which the Gaussian approximation is proposed. It seems reasonable, then, to choose the parameters of the Gaussian approximation in a manner that would result in the same $FWHM$ of projection on $x$ and $\theta$ as for the actual radiation Wigner function.

Clearly, the Gaussian approximation matched to the Wigner function at one point, should remain matched to it everywhere downstream in the drift.  This requirement is automatically satisfied for the Gaussian and non-Gaussian $FWHM_{\theta}$, which stay constant in a drift.  Far from the aperture, it is also satisfied for the $FWHM_x$, because in both cases $FWHM_x$ grows (asymptotically) linearly with distance, with the coefficient of proportionality found to be the (constant) $FWHM_{\theta}$. For a Gaussian beam, this is a well known fact that follows from writing the evolution equation for the $\Sigma$-matrix in the form 
\begin{equation}
\Sigma(s')  = M(s \to s') \Sigma(s) M^T(s \to s') \; ,
\end{equation}
where $M(s \to s')$ is the transport matrix in free space from the longitudinal position $s$ to $s'$.  From the $\sigma_{xx}$ component of this equation one finds immediately the asymptotically linear growth of the rms beam size at large distances $L$ from the waist, 
\begin{equation} 
\sigma_x(L) = \sqrt{\sigma_x^2(0) +\sigma_{\theta}^2 L^2} \to \sigma_{\theta}L \; 
\end{equation}
(assuming the waist location at $s = 0$).  Because for a Gaussian distribution 
\begin{equation}
\sigma = FWHM \; /2\sqrt{2\ln(2)} \; , 
\end{equation}
we have in terms of the FWHM parameters the asymptotic relation 
\begin{equation} 
FWHM_x \approx FWHM_{\theta} L \; .
\end{equation}

\begin{figure}[!htb]
   \centering
   \includegraphics*[width=1.0\columnwidth]{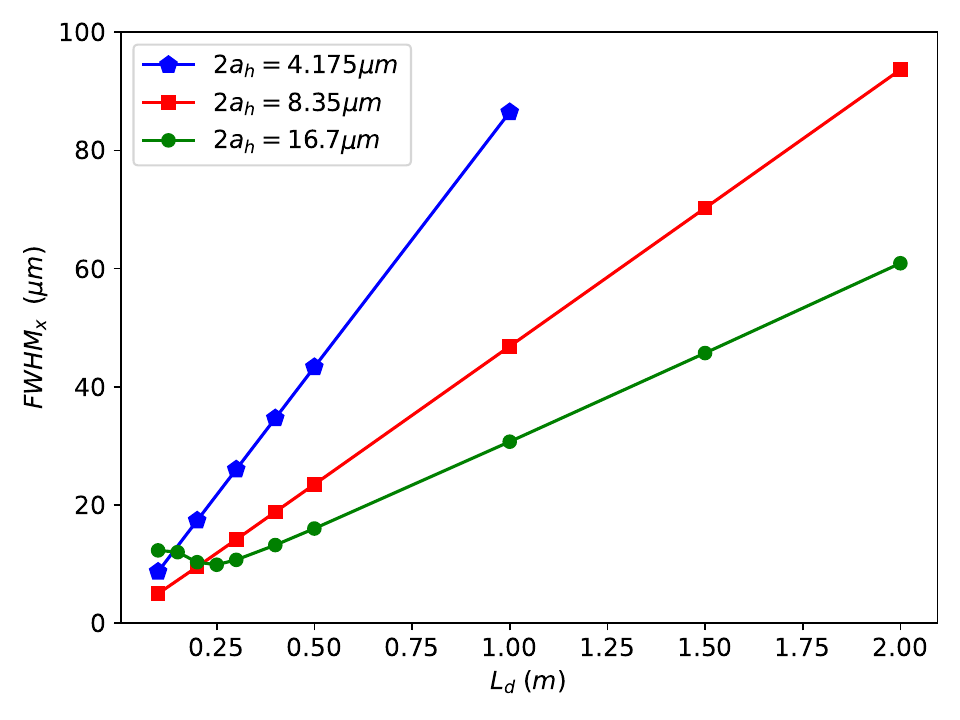}
   \caption{$FWHM$ of the Wigner function projection on the $x$ axis, as a function of distance $L_d$ from the aperture for a physical (Lorentzian) beam, for 3 values of the hard-edge aperture size $2a_h$.}
   \label{fig:size_vs_L}
\end{figure}

For the exact Wigner function, the asymptotic linear growth of the beam size $FWHM_x$ with rate $FWHM_{\theta}$ can be ascertained numerically. This is illustrated in Figure~\ref{fig:size_vs_L} for three different values of the hard-edge aperture size, $2 a_h = 16.7$, $8.35$, and $4.175 \mu$m, and the same values of the X-ray radiation parameters as in Fig.~\ref{fig:WF_10_30_cm_and_proj_on_x} ($\lambda = 3.98$ $10^{-10}$ m and $m^2 = 5$).  With decreasing aperture size, transition to the far-field regime occurs sooner, as may be expected from qualitative considerations. To the extent that the parameters of this numerical example are within the domain of practical interest, one may also conclude that the transition to the far-field regime occurs over distances small compared to the typical drift length in an X-ray beamline.

Thus, once the $FWHM_{\theta}$ for the Wigner function of the physical beam is known, a Gaussian approximation is effectively known, as well.  
Choosing $\sigma_{\theta\theta ,f}$ of the Gaussian approximation past the aperture so as to match the $FWHM_{\theta}$ of the approximation to $FWHM_{\theta}$ of the physical Wigner function, we can proceed to solve Eq.~\eqref{eq:sigma_tt_f}, a quadratic equation for the variable $a_g^2$, to determine the parameter $a_g$ of the effective Gaussian aperture that would result in the Gaussian Wigner function out of the aperture having the phase space footprint matching that of the physical Wigner function. In the special case of the incident beam with $\sigma_{x\theta} = 0$, which by analogy with particle beams may be called a matched beam, the aperture parameter is 
\begin{equation}
a_g =  \frac{\lambda}{4\pi}\sqrt{ \frac{2} 
{\sigma_{\theta\theta ,f} -\sigma_{\theta\theta}} } \; .
\end{equation}
In the general case with $\sigma_{x\theta}$ not necessarily zero, the solution is given by 
\begin{equation}
a_g = \sqrt{ \sqrt{b^2 +4c} -b } 
\end{equation} 
with 
\begin{equation}
b = \sigma_{xx} +\frac{\sigma_{x \theta}^2 -(\lambda /4\pi)^2}{\sigma_{\theta\theta ,f} -\sigma_{\theta\theta}} 
\end{equation} 
and 
\begin{equation}
c = \frac{\sigma_{xx}}{\sigma_{\theta\theta ,f} -\sigma_{\theta\theta}} 
\left( \frac{\lambda}{4\pi} \right)^2 \; .
\end{equation} 
The remaining elements of the $\Sigma$-matrix of the Gaussian approximation immediately after the aperture are then supplied by Eqs.~\eqref{eq:sigma_xx_f} and \eqref{eq:sigma_xt_f}. 

One question that requires further study is the dependence of the $a_g$ and the exact $FWHM_{\theta}$ on the system parameters such as $a_h$, $\lambda$, and the beam emittance $\epsilon$.  In the numerical example considered in this section, $a_g \approx 0.62 a_h$, but the exact dependence is nonlinear.  Likewise, the analytic dependence of $FWHM_{\theta}$ on $a_h$, $\lambda$ and $\epsilon$ that determines the far-field asymptotics is complicated and cannot be expressed in elementary functions.  We will address these questions in a separate publication.

\section{Software}
\label{sec:06}
The envelope algorithm we describe here may be integrated into an existing X-ray optics simulation code. We briefly describe this and report on the
implementation with the Sirepo graphical interface to the SHADOW ray tracing code~\cite{SirepoShadow}. 

X-ray optics software for modeling synchrotron radiation beamlines allows the user to describe the radiation source in terms of electron beam and magnetic
field properties and then to set up the beamline optical elements such as mirrors, crystals, gratings and other focusing and transmission elements. These codes exist with different approximations, two of the main ones being ray optics and wave optics. We here consider the ray optics code SHADOW. 

\begin{figure}[!htb]
   \centering 
   \includegraphics*[width=75 mm]{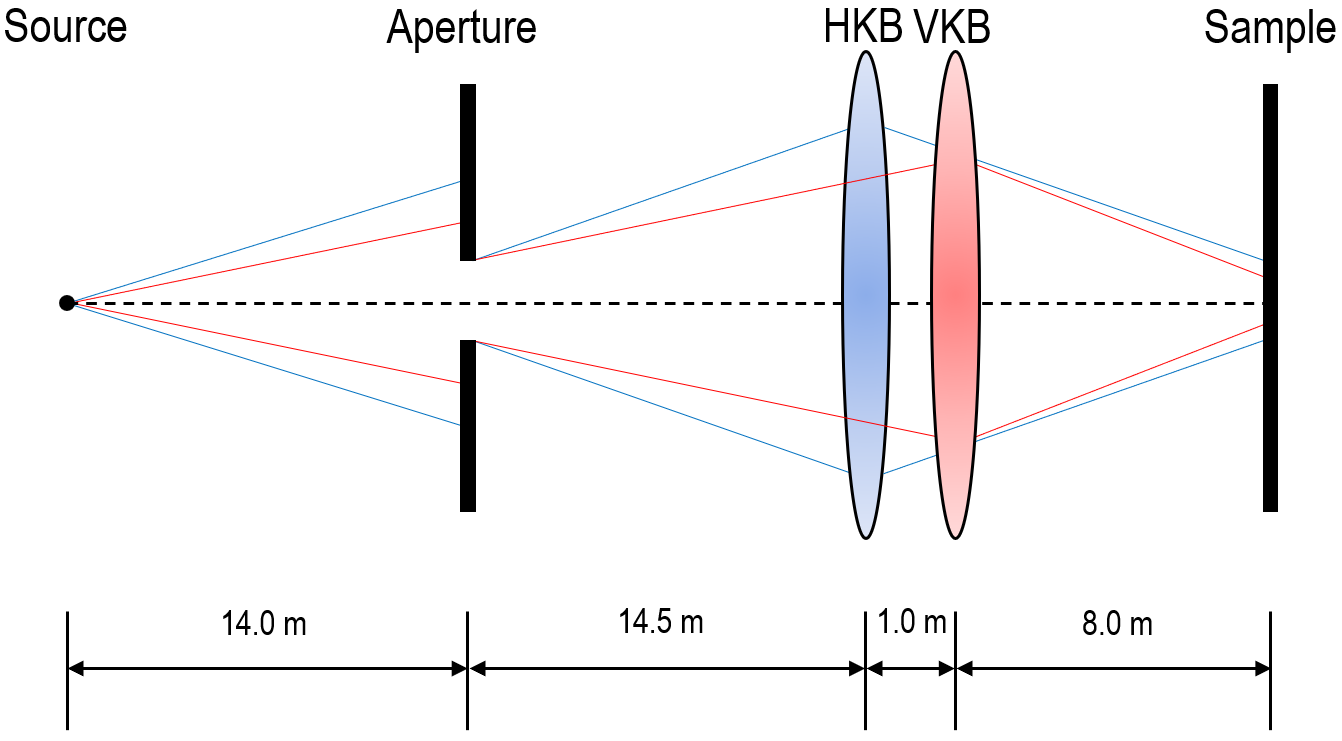}
   \caption{Diagram of KB-mirror beamline with beam defining aperture.  Blue and red lines represent horizontal and vertical beam projection respectively.  HKB and VKB are horizontally and vertically focusing elliptical mirrors represented by lenses in the diagram.}
   \label{fig:kb_aperture_beamline01}
\end{figure}

\begin{figure}[!htb]
   \centering
   \includegraphics*[width=75.0 mm]{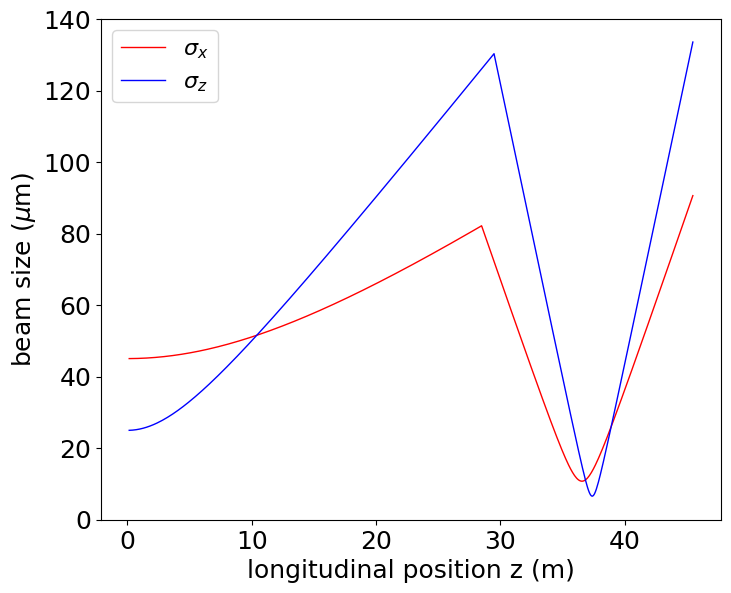}
   \caption{Horizontal and vertical X-ray beam size evolution through KB beamline without aperture.  The ray-tracing code Shadow was used for the transfer matrix calculations. A value of $m^2 = 1.1$ was used to set the initial conditions.}
   \label{fig:kb_no_aperture_size01}
\end{figure}

\begin{figure}[!htb]
   \centering
   \includegraphics*[width=75.0 mm]{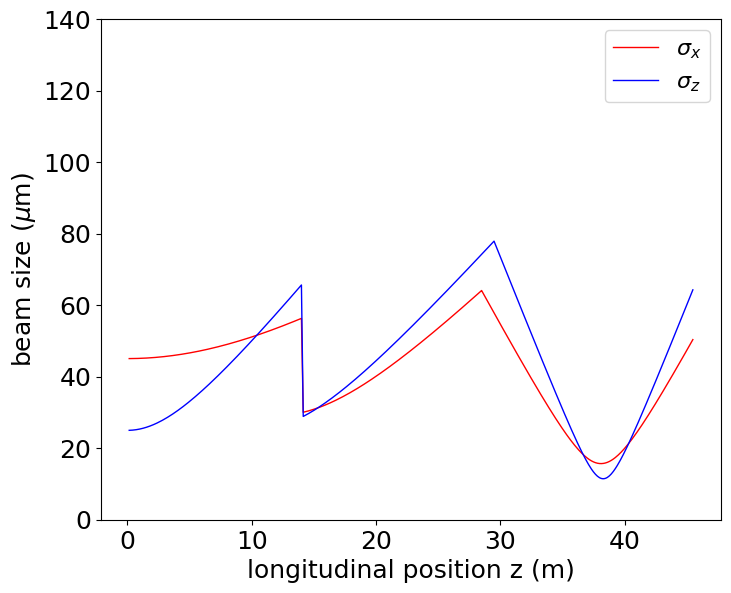}
   \caption{Horizontal and vertical X-ray beam size evolution through KB beamline with Gaussian aperture. A value of $m^2 = 1.1$ was used to set the initial conditions.}
   \label{fig:kb_single_aperture_size01}
\end{figure}

After setting up the optical beamline elements and their positions and orientation, the user creates an initial set of rays for tracing through the beamline. To compute the matrices between apertures, we track an initial central ray, along with 4 other rays, offset by a small value $\epsilon$ in each of the phase space dimensions $\vec r,\vec \theta$.
By tracking these 4 rays through all the optical elements up to the next aperture and relating that to the initial values, we construct the transfer matrices $M_j$ down the beamline. Next we identify the rectangular apertures and relate their sizes to an associated Gaussian aperture. Applying our formulae, we evolve the covariance matrix at position $s_j$, $\Sigma_j$
and the centroid. This is implemented as the ``beam statistics report" in Sirepo SHADOW. As an example, we work with a simple KB beamline with a single aperture as shown in Figure
\ref{fig:kb_aperture_beamline01}. The resulting beam sizes with and without the aperture are given in Figures \ref{fig:kb_no_aperture_size01} and \ref{fig:kb_single_aperture_size01}.

\begin{figure}[!htb]
   \centering
   \includegraphics*[width=75.0 mm]{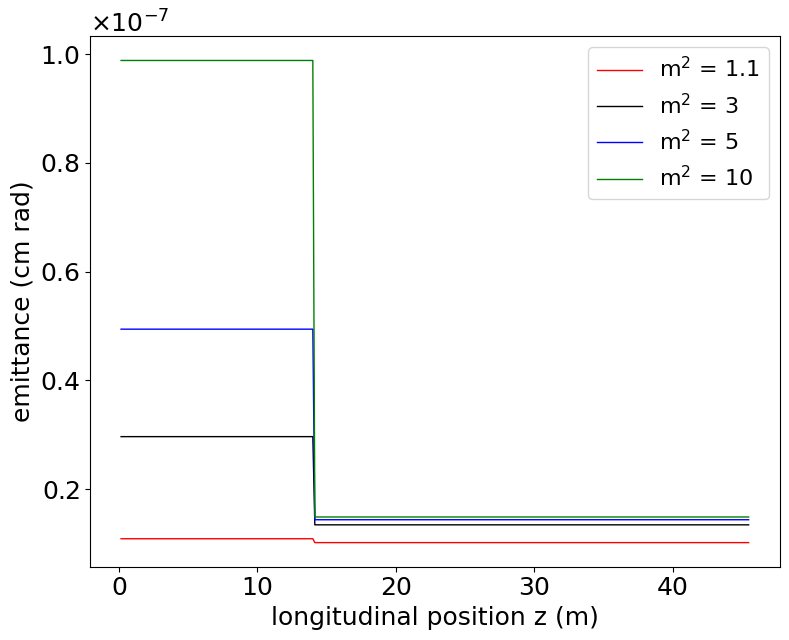}
   \caption{Evolution of horizontal X-ray beam emittance through KB-mirror aperture beamline with varying values of beam quality factor.}
   \label{fig:kb_emittance01}
\end{figure}

The evolution of the emittance is given in Figure \ref{fig:kb_emittance01}. The value of $m^2$ is varied, and we can see the impact of the aperture. We recall that emittance is conserved during the linear transport sections. The less coherent the beam is, the stronger the impact is of the aperture on the emittance. In the coherent case, the aperture doesn't change the emittance, as shown in Eq. \ref{emittance_change_aperture}.

\begin{figure}[!htb]
   \centering
   \includegraphics*[width=75.0 mm]{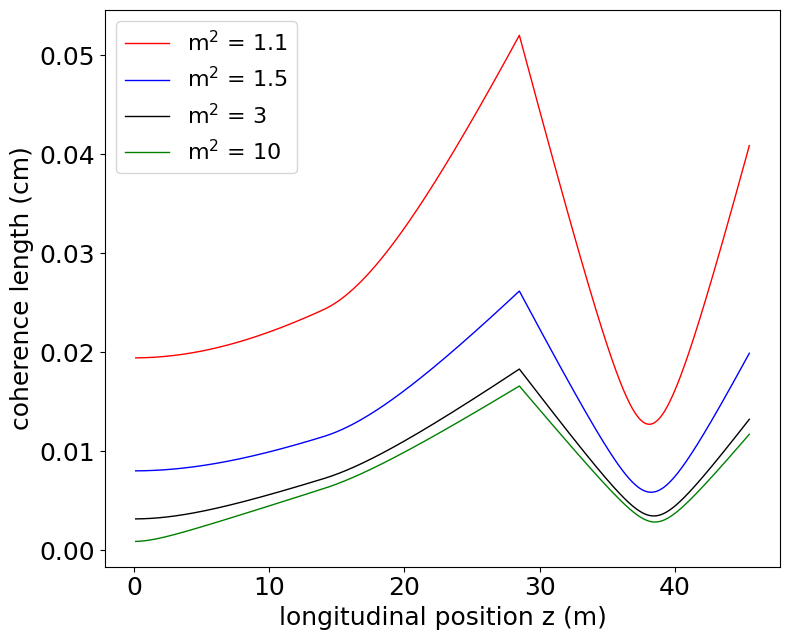}
   \caption{Evolution of horizontal coherence length through KB-mirror aperture beamline with varying values of beam quality factor.}
   \label{fig:kb_cohlength01}
\end{figure}

The coherence length evolution for varying values of $m^2$ is given in Figure \ref{fig:kb_cohlength01}.  Here, we observe the result described in 
Subsection \ref{coherence_length_through_aperture} that the coherence length does not change during passage through the aperture. The slope is seen to change, however, as expected, since it is proportional to the beam size.

\section{Summary and Discussion}
\label{sec:07}
We have discussed the use of Gaussian Wigner functions to describe the propagation of partially coherent radiation down optical beamlines. 
This framework, when applicable, is remarkably compact and convenient, represented by an evolving covariance matrix and phase space centroid coordinates. We described a simplified beam transport framework we call a  matrix-aperture beamline composed of linear transport sections and a series of apertures, expanded about the evolving beam centroid. The 
description of the radiation in terms of a Wigner function allows both fully and partially coherent radiation to be described. We outlined the one-to-one mapping between Gaussian Wigner functions and the more often used Gaussian Schell model cross-spectral density. Using this relationship, we derived an expression for the coherence length and other coherence properties in terms of the covariance matrix. We also remind the reader of the relationship between Gaussian distributions and the Twiss parameters and emittance from charged particle beam dynamics. This allows one to use a familiar notation and apply intuition from particle beam dynamics to the radiation propagation, particularly valuable in the important application of synchrotron radiation in which a Gaussian electron beam produces the partially coherent radiation after passage through a bending magnet or undulator.

The challenge with using the Gaussian Wigner function model for radiation transport arises with the use of rectangular (hard-edge) apertures in the beamline which break the Gaussian condition. We derived an exact expression for the Wigner function following a rectangular aperture. Although directly following the aperture, the intensity distribution is seen to be double peaked (not well represented by a Gaussian), looking a modest distance downstream, one finds, the single peak is recovered, making the Gaussian approximation a suitable choice. A further difficulty is discovered by examining the analytical expression following the hard-edge aperture. The intensity is seen to have a Lorentzian distribution even in the far field such that second moments are formally infinite. However, if one uses FWHM to define the covariance matrix, we find that a consistent model may be constructed with a direct mapping between the exact Wigner function and the Gaussian model. One may then use the exact hard-edge expression for guidance in finding an adequate Gaussian Wigner function model.

Next, we examined the case of a Gaussian aperture to approximate the hard-edge case. We derived expressions for the evolution of the covariance matrix and the centroid following passage through a given Gaussian aperture. For the separable 2D case, we found a number of convenient analytical expressions for how the beam sizes and divergences as well as derived quantities such as the beam emittance change during passage through the Gaussian aperture. Further, we analyzed how the coherence lengths evolve and found the surprising result that they are unchanged upon passage through the aperture. This is seen to be consistent with other coherence properties increasing due to the aperture. Particularly, as the beam size is reduced through the aperture, the unchanged coherence length in fact represents a relative increase in coherence; and evolving further down the beamline, it is seen that the coherence length will increase overall.

We performed numerical studies of the exact hard edge Wigner function to relate the hard edge size and the associated Gaussian aperture size that gives rise to the same covariance matrix. There is some complexity in this relationship and more detailed studies are needed to fully map out the relationship.

The algorithms described here have been implemented in the Sirepo interface to the ray optics software, SHADOW. The transport matrices for the linear sections are computed with ray tracing and one can plot the beam sizes, divergences and coherence lengths down the beamline for real synchrotron radiation beamline models. This adds a convenient missing tool for rapid understanding of the optical properties of different beamline designs. By coupling these calculations to beam diagnostics measurements, one may also construct online models analogously to how Twiss parameters are used to help understand and control linear and circular particle beam transport systems, filling in an important gap particularly on X-ray beamlines.

\begin{acknowledgments} 
This work was supported by the U.S. Department of Energy, Office of Science, Office of Basic Energy Sciences, under Award Number DE-SC0020593.
\end{acknowledgments}

\begin{appendix}

\section{2D Gaussian Wigner function for a Non-Matched Beam with Misalignments in Angle and Position Crossing a Hard-Edge Aperture}
\label{app:A}
As described in Section~\ref{sec:03}, the radiation Wigner function after a hard-edge aperture is calculated via a convolution in $\theta$ of the radiation Wigner function before the aperture with the formally defined ``Wigner function'' of the aperture.  To compute the convolution, we employ the convolution theorem:
\begin{equation}
f \ast g = IFT[FT[f] \cdot FT[g]] \; ,
\end{equation}
where $FT$ and $IFT$ are, respectively, the Fourier transform and the inverse Fourier transform, defined as in Eqs.~\eqref{eq:ft} and~\eqref{eq:ift}. 

In order to reduce clutter in the equations that follow, we introduce at this point some shorthand notation:
\begin{eqnarray*}
&\tilde{a} \equiv \frac{\beta}{2\epsilon} \\
&\tilde{\theta} \equiv \theta_0 -\frac{\alpha}{\beta}(x -x_0) \\
&c \equiv \frac{\sqrt{\tilde{a}}}{\pi} = \frac{1}{\pi} \sqrt{ \frac{\beta}{2\epsilon} } \\
&\kappa \equiv \frac{4(a_h -|x|)}{\lambda} \\
&L_a \equiv -\frac{2(a_h -|x|)}{\lambda} = -\frac{\kappa}{2} \\
&L_b \equiv \frac{2(a_h -|x|)}{\lambda} = \frac{\kappa}{2}\\
&\mu \equiv i\frac{\tilde{a}}{\pi}(\theta -\theta_0) \; .
\end{eqnarray*}
With this notation, the Wigner function of the incident radiation is given by 
\begin{equation}
W_i(x, \theta) =\frac{1}{2\pi\epsilon} \exp \left(  -\frac{(x-x_0)^2}{2\epsilon\beta}  \right) \exp \left( -\tilde{a} (\theta -\theta_0)^2   \right) \; .
\end{equation}
The $\theta$-dependent inputs to computing the convolution via the convolution theorem are 
\begin{equation}
FT[W_a(x, \theta)] = FT\left[ \frac{\sin(\kappa \pi\theta)}{\pi\theta} \right] = \Pi_{\kappa/2} (t) 
\end{equation}
with 
\begin{equation}
\Pi_a(x) = 
\left\{ \begin{aligned} 
   &1,  \;\; |x| < a  \; ,\\
  & 1/2 \; , \; |x| = a > 0 \; , \\
  &0 \; , |x| > a \; ,
\end{aligned} \right.
\end{equation}
and
\begin{equation}
\begin{split}
&FT[\exp \left( -\tilde{a} (\theta -\theta_0)^2 \right)] = e^{-i 2\pi \tilde{\theta} t} FT[\exp(-\tilde{a}\theta^2)] \\
&= e^{-i 2\pi \tilde{\theta} t} \sqrt{\frac{\pi}{\tilde{a}}}\exp\left( - \frac{\pi^2 t^2}{\tilde{a}} \right) \\
&= \exp(-\tilde{a}\tilde{\theta}^2) \frac{1}{\sqrt{\tilde{a}/\pi}} \exp \left( - \frac{(t +i\tilde{\theta}\tilde{a}/\pi)^2}{\sqrt{\tilde{a}}/\pi} \right) \; .
\end{split}
\end{equation}
Now, using the shorthand notation introduced above, 
\begin{widetext}
\begin{equation}
\begin{split}
IFT &\left[ \Pi_{\kappa/2} (t) \frac{1}{\sqrt{\tilde{a}/\pi}} \exp \left( - \frac{(t +i\tilde{\theta}\tilde{a}/\pi)^2}{\sqrt{\tilde{a}}/\pi} \right)  \right]  
= \frac{1}{c\sqrt{\pi}} \int^{L_b}_{L_a} \exp \left[-\frac{(t +i\tilde{\theta}\tilde{a}/\pi)^2}{c^2}  \right] \exp(i 2\pi \theta t) dt \\
&= \exp( -\tilde{a}\theta (\theta -2\tilde{\theta}) ) 
\frac{1}{c\sqrt{\pi}} \int^{L_b}_{L_a} \exp \left[ -\frac{(t -i \pi c^2 (\theta -\tilde{\theta}))^2}{c^2}   \right] dt  \\
&
=  \exp( -\tilde{a}\theta (\theta -2\tilde{\theta}) ) 
\frac{1}{2} \left[ \erf \left( \frac{L_b -\mu}{c} \right)  - \erf \left( \frac{L_a -\mu}{c} \right) \right] \; ,
\end{split}
\end{equation}
where we make use of a well-known relation for the error function, 
\begin{equation}
\begin{split}
&\frac{1}{c\sqrt{\pi}} \int_{L_a}^{L_b}  \exp \left(  -\frac{(x-\mu)^2}{c^2}  \right) dx 
= \frac{1}{2} \left[ \erf \left( \frac{L_b -\mu}{c} \right)   -\erf \left( \frac{L_a -\mu}{c} \right)  \right] \; .
\end{split}
\end{equation}
\end{widetext}

Using the relations
\begin{equation}
\erf(-z) = -\erf(z), \;\;\; \erf(\bar{z}) = \overline{\erf(z)} \; ,
\end{equation}
we can finally write the expression for the radiation Wigner function after the hard-edge aperture as 
\begin{equation}
\begin{split}
&W_f(x, \theta) = W_i(x, \theta)\ \Pi_{a_h}(x) \\ 
& \times \operatorname{Re} \left\{ \erf \left[ \sqrt{\frac{2\epsilon}{\beta}} \left( \frac{2\pi (a_h -|x|)}{\lambda}   \right. \right. \right. \\
& \left. \left. \left. +i \left( \frac{\beta}{2\epsilon}(\theta -\theta_0) +\frac{\alpha}{2\epsilon}(x-x_0)   \right)   \right) \right] \right\}  \; .
\end{split}
\end{equation}

\section{2D Gaussian Wigner function for a Beam without Misalignments after a Gaussian Aperture}
\label{app:B}

The Wigner function of the radiation after the aperture, $W_f(x, \theta)$,  is computed via the convolution in $\theta$ 
of the radiation Wigner function before the aperture, $W_i(x, \theta)$, in the form of the last line of Eq.~\eqref{w_i_2D}, and the ``Wigner function'' of the aperture, $W_a(x, \theta; a_g)$, given by Eq.~\eqref{eq:apertW}.  Factoring out the parts of the two Wigner functions that depend on $\theta$, both of which are Gaussian in $\theta$, and using the fact that the convolution of two Gaussian distributions is again a Gaussian whose mean and the covariance matrix are respectively the sums of the means and covariance matrices of the input distributions, we find:
\begin{equation}
\begin{split}
\frac{1}{\sqrt{2\pi} \sqrt{\epsilon / \beta} } \exp \left[  -\frac{\beta}{2\epsilon}  \left( \theta +\frac{\alpha}{\beta}x \right)^2  \right] \\
\ast \frac{1}{\sqrt{2\pi} \lambda/\sqrt{8}\pi a_g} \exp \left[   -\frac{\theta^2}{2(\lambda /\sqrt{8} \pi a_g)^2} \right] \\
= \frac{1}{\sqrt{2\pi} \sigma_{\theta,f}} \exp \left[  -\frac{1}{2\sigma_{\theta,f}^2}  \left( \theta +\frac{\alpha}{\beta}x \right)^2  \right]  \; ,
\end{split}
\end{equation}
where we have introduced a shorthand notation 
\begin{equation}
\sigma_{\theta,f}^2 = \frac{\epsilon}{\beta} +\frac{\lambda^2}{8\pi^2 a_g^2}  = \frac{\det\Sigma}{\sigma_{xx}} +\frac{\lambda^2}{8\pi^2 a_g^2} \; .
\label{eq:sigma_theta_f}
\end{equation}
Here as well as throughout the paper the subscript ``$f$'' denotes quantities after the aperture, and the absence of a subscript indicates a quantity before the aperture. 

The radiation Wigner function out of the aperture is then 
\begin{equation}
\begin{split}
&W_f(x, \theta) \\
&\propto \exp{ \left[  -\left( \frac{1}{a_g^2} +\frac{1}{2\sigma_{xx}} \right) x^2 \right.
\left. -\frac{1}{2\sigma_{\theta,f}^2} \left( \theta -\frac{\sigma_{x\theta}}{\sigma_{xx}}x \right)^2  \right] }\\
&= \exp \left[  -\frac{1}{2} \left( \left( \frac{2}{a_g^2} +\frac{1}{\sigma_{xx}} +\frac{\sigma_{x\theta}^2}{\sigma_{\theta,f}^2 \sigma_{xx}^2} \right) x^2 \right. \right.\\
&\left. \left. -\frac{2\sigma_{x\theta}}{\sigma_{\theta,f}^2 \sigma_{xx}} x\theta +\frac{1}{\sigma_{\theta,f}^2} \theta^2 \right)  \right]  \; .
\end{split}
\label{w_f_exponent}
\end{equation}
Viewing Eq.~\eqref{w_f_exponent} as having the functional form of Eq.~\eqref{w_i_2D} (subject furthermore to the identity constraint Eq.~\eqref{twiss_ident}), we find the emittance $\epsilon_f$ and the elements of the $\Sigma_f$ matrix after the aperture in terms of the aperture size parameter and the elements of the $\Sigma$ matrix before the aperture: 
\begin{equation}
\begin{split}
&\epsilon_f^2 = \sigma_{\theta,f}^2 \left( \frac{2}{a_g^2} +\frac{1}{\sigma_{xx}} \right)^{-1} \\
&= \left( \frac{\epsilon^2}{\sigma_{xx}} +\frac{\lambda^2}{8\pi^2 a_g^2} \right) \left( \frac{2}{a_g^2} +\frac{1}{\sigma_{xx}} \right)^{-1} \; ,
\end{split}
\label{eq:eps2_f} 
\end{equation}
\begin{equation}
\sigma_{xx,f} =  \left( \frac{2}{a_g^2} +\frac{1}{\sigma_{xx}} \right)^{-1}  = \frac{a_g^2\sigma_{xx}}{a_g^2 +2\sigma_{xx}}\; , 
\label{eq:sigma_xx_f_apx}
\end{equation} 
\begin{equation}
\sigma_{x\theta,f} =  \frac{\sigma_{x\theta}}{\sigma_{xx}} \left( \frac{2}{a_g^2} +\frac{1}{\sigma_{xx}} \right)^{-1}  = \frac{a_g^2\sigma_{x\theta}}{a_g^2 +2\sigma_{xx}}\; , 
\label{eq:sigma_xt_f_apx}
\end{equation}
\begin{equation}
\begin{split}
\sigma_{\theta\theta,f} = \sigma_{\theta,f}^2 +  \frac{\sigma_{x\theta}^2}{\sigma_{xx}^2} \left( \frac{2}{a_g^2} +\frac{1}{\sigma_{xx}} \right)^{-1} \\
= \sigma_{\theta\theta} -\frac{2\sigma_{x\theta}^2}{a_g^2 +2\sigma_{xx}}  +\frac{\lambda^2}{8 \pi^2 a_g^2}\; .
\end{split}
\label{eq:sigma_tt_f_apx}
\end{equation}

\section{2D Gaussian Wigner function for a Beam with Centroid Offsets in Angle and Position after a Gaussian Aperture}
\label{app:B_w_offsets}
Denoting the elements of the inverse covariance matrix after the aperture by
\begin{equation} 
\Sigma_f^{-1} = 
\begin{pmatrix}
    s_{11} & s_{12}\\
    s_{12} & s_{22}
\end{pmatrix} \; ,
\end{equation} 
and the misalignments in position and angle before and after the aperture respectively by $x_0$, $\theta_0$, $x_{0,f}$, and $\theta_{0,f}$, we can write the polynomial in the exponent of the radiation Wigner function after the aperture as 
\begin{equation}
\begin{split}
-\frac{1}{2}(x -x_{0,f}, \theta -\theta_{0,f}) \Sigma_f^{-1} (x -x_{0,f}, \theta -\theta_{0,f})^T = \\
= -\frac{s_{11}}{2}x^2  -s_{12}x\theta -\frac{s_{22}}{2}{\theta}^2 \\
+(s_{11}x_{0,f} +s_{12}\theta_{0,f}) x +(s_{12}x_{0,f} +s_{22}\theta_{0,f})\theta \\
-\frac{s_{11}}{2}x_{0,f}^2  -s_{12}x_{0,f}\theta_{0,f} -\frac{s_{22}}{2}\theta_{0,f}^2 \; .
\end{split}
\end{equation}
The coefficients by non-zero powers of the phase space variables in this polynomial we now equate with the coefficients by same powers in the second degree polynomial in the exponential part of the post-aperture Wigner function obtained by convolution in $\theta$ of the radiation Wigner function before the aperture and the ``Wigner function'' of the aperture,
\begin{equation}
\begin{split}
&W(x, \theta) \propto \exp \left(  -\frac{(x-x_0)^2}{2\epsilon\beta} -\frac{x^2}{a_g^2} \right) \\
&\times    \exp \left(  -\frac{1}{2(\epsilon/\beta)}  \left( \theta -\theta_0+\frac{\alpha}{\beta}(x -x_0)  \right)^2  \right) \\
&\ast_{\theta}  \exp \left[   -\frac{\theta^2}{(\lambda /2 \pi a_g)^2} \right] \\
&\propto \exp \left(  -\frac{(x-x_0)^2}{2\epsilon\beta} -\frac{x^2}{a_g^2} \right)  \\
&\exp \left[  -\frac{1}{2\sigma_{\theta,f}^2}  \left( \theta -\theta_0+\frac{\alpha}{\beta}(x -x_0) \right)^2  \right] \; ,
\end{split}
\end{equation}
where $\sigma_{\theta,f}$ is given by the same Eq.~\eqref{eq:sigma_theta_f} as in the case of no misalignment.  Moreover, for the coefficients by the $x^2$, $x\theta$, and $\theta^2$, we obtain 
\begin{equation}
\begin{split}
s_{11} = \frac{1}{\sigma_{xx}} +\frac{2}{a_g^2} +\frac{1}{\sigma_{\theta,f}^2}\frac{\sigma_{x\theta}^2}{\sigma_{xx}^2} \; , \\
s_{12} = \frac{-1}{\sigma_{\theta,f}^2}\frac{\sigma_{x\theta}}{\sigma_{xx}} \; , \;\;\;  s_{22} = \frac{1}{\sigma_{\theta,f}^2} \; ,
\end{split}
\end{equation}
which are the same expressions as in the case of no misalignments, so that the elements of the $\Sigma_f$ matrix after the aperture, 
\begin{equation}
\begin{split}
\sigma_{xx, f} = \frac{s_{22}}{s_{11}s_{22} -s_{12}^2}, \;\; \sigma_{x\theta, f} = \frac{-s_{12}}{s_{11}s_{22} -s_{12}^2}, \\ \sigma_{\theta\theta, f} = \frac{s_{11}}{s_{11}s_{22} -s_{12}^2},
\end{split}
\end{equation}
are related to the elements of the $\Sigma$ matrix before the aperture in the same way as in the case of no misalignments, Eqs.~\eqref{eq:sigma_xx_f}-\eqref{eq:sigma_tt_f}. 
To relate the beam misalignments in position and angle after the aperture to those before the aperture, we equate the coefficients by $x$ and $\theta$:
\begin{equation}
s_{11}x_{0,f} +s_{12}\theta_{0,f} = \left( s_{11} -\frac{2}{a_g^2}   \right) x_0 +s_{12}\theta_0 \; ,
\end{equation}
\begin{equation}
s_{12}x_{0,f} +s_{22}\theta_{0,f} = s_{12} x_0 +s_{22}\theta_0 \; ,
\end{equation}
whence 
\begin{equation}
x_{0,f} = x_0 -\sigma_{xx,f} \frac{2}{a_g^2} x_0 = \frac{\sigma_{xx,f}}{\sigma_{xx}} x_0
\label{eq:x_offset_f_apx}
\end{equation}
and
\begin{equation}
\theta_{0,f} = \theta_0 -\frac{2}{a_g^2} \sigma_{x\theta, f} x_0 \; .
\end{equation}

\section{4D Gaussian Wigner function for a Beam without Centroid Offsets after a Gaussian Aperture}
\label{app:C}
It is possible to extend the construction of an effective Gaussian aperture presented above from $2D$ to $4D$, without any separability assumptions.  Ordering the phase space variables as 
\begin{equation}
\vec{z} = 
\begin{pmatrix}
    \vec{r} \\
    \vec{\theta} 
\end{pmatrix} \; ,
\end{equation} 
where 
\begin{equation}
\vec{r} = 
\begin{pmatrix}
    x \\
    y  
\end{pmatrix} \; 
\end{equation} 
and 
\begin{equation}
\vec{\theta} = 
\begin{pmatrix}
    \theta_x \\
    \theta_y 
\end{pmatrix} \; ,
\end{equation} 
we write the inverse of the correlation matrix $\Sigma$ in terms of four $2\times2$ matrices $A$, $B$, $C$, and $D$: 
\begin{equation} 
\Sigma^{-1} = 
\begin{pmatrix}
    A & B\\
    C & D
\end{pmatrix} \; .
\end{equation} 
Because $\Sigma$ is symmetric, we have $\Sigma^{-1} = (\Sigma^{-1})^T$, and therefore $A = A^T$, $D = D^T$, and $B = C^T$. It follows that 
\begin{equation}
-\frac{1}{2} {\vec{z}}^T \Sigma^{-1} \vec{z} =  -\frac{1}{2} \left(  {\vec{\theta}}^T D \vec{\theta}  +2(C \vec{r})^T \vec{\theta}  +{\vec{r}}^T A \vec{r} \right)  \; .
\label{quadform} 
\end{equation}
For the purpose of computing the convolution in $\vec{\theta}$ when traversing an aperture, we re-arrange the r.h.s. of Eq.~\eqref{quadform} as 
\begin{equation}
\begin{split} 
-\frac{1}{2} {\vec{z}}^T \Sigma^{-1} \vec{z} =  -\frac{1}{2} \left(  (\vec{\theta} +D^{-1}C\vec{r} )^T D (\vec{\theta} +D^{-1}C \vec{r} ) \right. \\
\left. +{\vec{r}}^T (A -BD^{-1}C) \vec{r} \right)  \; .
\end{split} 
\label{eq:exponent_4D_no_offsets}
\end{equation}
The angular part of the (separable) ``Wigner function'' for the Gaussian aperture is likewise a Gaussian, which we write as 
\begin{equation}
W_{a,\theta} \propto \exp \left(  -\frac{1}{2} \vec{\theta}^T D_A \vec{\theta}  \right)  \; .
\end{equation}
When two multi-dimensional Gaussians are convolved in \textit{all} dimensions, the result is a Gaussian with the (vector) mean equal to the sum of the mean values of the two distributions, and the covariance matrix the sum of the covariance matrices of the convolved distributions.  For the convolution in angle sub-space of the Wigner functions of the incident wave and the aperture, we find therefore 
\begin{widetext}
\begin{equation}
\begin{split} 
\exp \left(    -\frac{1}{2}  (\vec{\theta} +D^{-1}C\vec{r} )^T D (\vec{\theta} +D^{-1}C \vec{r} ) \right) * \exp \left(  -\frac{1}{2} \vec{\theta}^T D_A \vec{\theta}  \right) \\
\propto    \exp \left(    -\frac{1}{2}  (\vec{\theta} +D^{-1}C\vec{r} )^T (D^{-1} + D_A^{-1})^{-1} (\vec{\theta} +D^{-1}C \vec{r} ) \right) .
\end{split} 
\end{equation}
\end{widetext}
The constituent $2\times2$ blocks of the $4\times4$ \textit{inverse} of the covariance matrix after the aperture are thus given by
\begin{equation}
A_f = A +A_A -BD^{-1}C +BD^{-1} (D^{-1} +D_A^{-1})^{-1} D^{-1}C   \; ,
\label{eq:A_f}
\end{equation}
\begin{equation}
B_f = BD^{-1} (D^{-1} +D_A^{-1})^{-1}  \; ,
\end{equation} 
\begin{equation}
C_f = (D^{-1} +D_A^{-1})^{-1} D^{-1}C = B_f^T  \; ,
\end{equation}
and 
\begin{equation}
D_f = (D^{-1} +D_A^{-1})^{-1}  \; ,
\label{eq:D_f}
\end{equation}
where, in terms of the parameters $a_x$ and $a_y$ specifying the Gaussian aperture in the $x$ and $y$ directions, 
\begin{equation}
A_A = 
 \begin{pmatrix}
    2/a_x^2 & 0\\
    0 & 2/a_y^2
\end{pmatrix} \; 
\label{eq:A_A}
\end{equation} 
and
\begin{equation}
D_A = \frac{8\pi^2}{\lambda^2} 
 \begin{pmatrix}
    a_x^2 & 0\\
    0 & a_y^2
\end{pmatrix} \; ,
\label{eq:D_A}
\end{equation} 
hence 
\begin{equation}
D_A^{-1} = \frac{\lambda^2}{8\pi^2} 
 \begin{pmatrix}
    1/a_x^2 & 0\\
    0 & 1/a_y^2
\end{pmatrix} \; .
\end{equation}

The covariance matrix $\Sigma_f$ after the aperture is then computed via a numerical inversion procedure. 

\section{4D-Phase-Space Gaussian Wigner function for a Beam with Centroid Offsets in Angle and Position after a Gaussian Aperture}
\label{app:D}
The analysis in the case of coupled $4D$ Gaussian Wigner function passing through a Gaussian aperture parallels that in the $2D$ case.  We retain the notation from the previous section, adding the subscripts ``$0$'' and ``$0,f$'' to the $2D$ and $4D$ phase-space-coordinate vectors to indicate the offsets before and after the aperture, respectively. The expressions for the $4D$ ``Wigner function'' of the aperture are the same as before, while the polynomial in the exponent of the radiation Wigner function, given in the absence of offsets by Eq.~\eqref{eq:exponent_4D_no_offsets}, becomes 
\begin{widetext}
\begin{equation}
\begin{split} 
&-\frac{1}{2} (\vec{z} -\vec{z}_0)^T \Sigma^{-1} (\vec{z} -\vec{z}_0)\\
&=  -\frac{1}{2} \left(  (\vec{\theta} -\vec{\theta}_0 +D^{-1}C (\vec{r} -\vec{r}_0) )^T D (\vec{\theta} -\vec{\theta}_0 +D^{-1}C (\vec{r} -\vec{r}_0) ) \right. 
\left. +(\vec{r} -\vec{r}_0)^T (A -BD^{-1}C) (\vec{r} -\vec{r}_0) \right)  \; .
\end{split} 
\label{eq:exponent_4D_with_offsets}
\end{equation}
\end{widetext}
To compute the radiation Wigner function after the aperture, we perform the convolution-in-$\theta$ of the radiation Wigner function and the aperture ``Wigner function'', the result of convolving two Gaussian functions being again a Gaussian whose mean and the covariance matrix is the sum of the means and covariance matrices of the two distributions being convolved. Leaving out a constant pre-factor, the result is 
\begin{widetext}
\begin{equation}
\begin{split} 
&W_f(\vec{r}, \vec{\theta}; \vec{r}_0, \vec{\theta}_0) 
\propto \exp \left[ -\frac{1}{2} \left( (\vec{r} -\vec{r}_0)^T (A +A_A -BD^{-1}C) (\vec{r} -\vec{r}_0)  \right. \right. \\
&+(\vec{\theta} -\vec{\theta}_0 +D^{-1}C (\vec{r} -\vec{r}_0) )^T 
\left. \left. (D^{-1} +D_A^{-1} )^{-1} (\vec{\theta} -\vec{\theta}_0 +D^{-1}C (\vec{r} -\vec{r}_0) ) \right) \right] \; .
\end{split} 
\label{eq:GWF_after_ap_from_conv}
\end{equation}
\end{widetext}
As in the previous section, we find the relation between the offsets and elements of $\Sigma$ before the aperture and the offsets and elements of $\Sigma_f$ after the aperture by matching the ``same-power'' in $\vec{r}$ and $\vec{\theta}$ terms in the exponent of Eq.~\eqref{eq:GWF_after_ap_from_conv} and in the exponent of the general expression for the Gaussian Wigner function after the aperture,
\begin{widetext}
\begin{equation}
-\frac{1}{2} (\vec{z} -\vec{z}_{0,f})^T \Sigma_f^{-1} (\vec{z} -\vec{z}_{0,f}) 
= -\frac{1}{2} \left( (\vec{r} -\vec{r}_{0,f})^T A_f (\vec{r} -\vec{r}_{0,f})  \right. 
\left. +2(C_f (\vec{r} -\vec{r}_{0,f}))^T (\vec{\theta} -\vec{\theta}_{0,f})  +(\vec{\theta} -\vec{\theta}_{0,f})^T D_f (\vec{\theta} -\vec{\theta}_{0,f}) \right) \; .
\label{quadform_f} 
\end{equation}
\end{widetext}
One finds immediately by examination of the second-order terms that the elements of the covariance matrix before and after the aperture are related in the same way as when there is no misalignment, \textit{i.e.}, by Eqs.~\eqref{eq:A_f}-\eqref{eq:D_f} with the auxiliary Eqs.~\eqref{eq:A_A}-\eqref{eq:D_A}. To find how the offsets before and after the aperture are related, we have to solve a system of two linear equations.  The first comes from equating the terms linear in $\vec{r}$, and the second is obtained by equating the terms linear in $\vec{\theta}$:
\begin{equation}
\begin{cases}
\vec{r}^T \left( A_f \vec{r}_{0,f} +B_f \vec{\theta}_{0,f} \right) = \vec{r}^T \left( (A_f -A_A) \vec{r}_0 +B_f \vec{\theta}_0 \right) \\
\vec{\theta}^T \left( C_f \vec{r}_{0,f} +D_f \vec{\theta}_{0,f} \right) = \vec{\theta}^T  \left( C_f \vec{r}_0 +D_f \vec{\theta}_0  \right)  
\end{cases},
\end{equation}
that is,
\begin{equation}
\begin{cases}
A_f (\vec{r}_{0,f} -\vec{r}_0) +B_f (\vec{\theta}_{0,f} -\vec{\theta}_0) =  -A_A \vec{r}_0  \\
C_f (\vec{r}_{0,f} -\vec{r}_0) +D_f (\vec{\theta}_{0,f} -\vec{\theta}_0)  =  \vec{0}   
\end{cases} \; ,
\end{equation}
or,
\begin{equation}
\Sigma_f^{-1} (\vec{z}_{0,f} -\vec{z}_0) = 
\begin{pmatrix}
    -A_A\vec{r}_0 \\
    \vec{0} 
\end{pmatrix} \; ,
\end{equation}
so that 
\begin{equation}
\vec{z}_{0,f} = \vec{z}_0 -\Sigma_f 
\begin{pmatrix}
    A_A\vec{r}_0 \\
    \vec{0} 
\end{pmatrix} \; .
\end{equation}
Writing out $\Sigma_f$ and $A_A$ matrices in terms of their elements, we see that this more general result is in agreement with the result for the $2D$ uncoupled case, as of course it should be.

\end{appendix}

\bibliography{PRABGaussian}

\end{document}